\documentclass[%
reprint,
superscriptaddress,
 amsmath,amssymb,
 aps,
]{revtex4-2}
\usepackage[utf8]{inputenc}

\usepackage{natbib}
\usepackage{graphicx}
\usepackage{color}
\usepackage{amssymb}
\usepackage{amsmath}
\usepackage{dcolumn}
\usepackage{ulem}
\newcommand{\angstrom}{\textup{\AA}}
\makeatletter
\newcolumntype{B}[3]{>{\boldmath\DC@{#1}{#2}{#3}}c<{\DC@end}}
\makeatother

\begin{document}

\title{Towards Universal Neural Network Potential for Material Discovery Applicable to Arbitrary Combination of 45 Elements}


\author{So Takamoto}
\thanks{Corresponding author}
\email{takamoto@preferred.jp}
\affiliation{Preferred Networks, Inc.}
\author{Chikashi Shinagawa}
\affiliation{Preferred Networks, Inc.}
\author{Daisuke Motoki}
\affiliation{Preferred Networks, Inc.}
\author{Kosuke Nakago}
\affiliation{Preferred Networks, Inc.}
\author{Wenwen Li}
\affiliation{Preferred Networks, Inc.}
\author{Iori Kurata}
\affiliation{Preferred Networks, Inc.}
\author{Taku Watanabe}
\affiliation{ENEOS Corporation, Central Technical Research Laboratory}
\author{Yoshihiro Yayama}
\affiliation{ENEOS Corporation, Central Technical Research Laboratory}
\author{Hiroki Iriguchi}
\affiliation{ENEOS Corporation, Central Technical Research Laboratory}
\author{Yusuke Asano}
\affiliation{ENEOS Corporation, Central Technical Research Laboratory}
\author{Tasuku Onodera}
\affiliation{ENEOS Corporation, Central Technical Research Laboratory}
\author{Takafumi Ishii}
\affiliation{ENEOS Corporation, Central Technical Research Laboratory}
\author{Takao Kudo}
\affiliation{ENEOS Corporation, Central Technical Research Laboratory}
\author{Hideki Ono}
\affiliation{ENEOS Corporation, Central Technical Research Laboratory}
\author{Ryohto Sawada}
\affiliation{Preferred Networks, Inc.}
\author{Ryuichiro Ishitani}
\affiliation{Preferred Networks, Inc.}
\author{Marc Ong}
\affiliation{Preferred Networks, Inc.}
\author{Taiki Yamaguchi}
\affiliation{Preferred Networks, Inc.}
\author{Toshiki Kataoka}
\affiliation{Preferred Networks, Inc.}
\author{Akihide Hayashi}
\affiliation{Preferred Networks, Inc.}
\author{Nontawat Charoenphakdee}
\affiliation{Preferred Networks, Inc.}
\author{Takeshi Ibuka}
\thanks{Corresponding author}
\email{ibuka.takeshi@eneos.com}
\affiliation{ENEOS Corporation, Central Technical Research Laboratory}
\date{\today}

\begin{abstract}
Computational material discovery is under intense study owing to its ability to explore the vast space of chemical systems.
Neural network potentials (NNPs) have been shown to be particularly effective in conducting atomistic simulations for such purposes.
However, existing NNPs are generally designed for narrow target materials, making them unsuitable for broader applications in material discovery.
To overcome this issue, we have developed a universal NNP called PreFerred Potential (PFP), which is able to handle any combination of 45 elements.
Particular emphasis is placed on the datasets, which include a diverse set of virtual structures used to attain the universality.
We demonstrated the applicability of PFP in selected domains: lithium diffusion in LiFeSO${}_4$F, molecular adsorption in metal-organic frameworks, an order--disorder transition of Cu-Au alloys, and material discovery for a Fischer--Tropsch catalyst. They showcase the power of PFP, and this technology provides a highly useful tool for material discovery. 

\end{abstract}
\maketitle
\section{Introduction}
Finding new and useful materials is a difficult task.
Because the number of possible material combinations in the real world is astronomically large \cite{chemicalspace}, methods for material exploration depending only on computer simulations are required to search through a vast number of candidate materials within a feasible amount of time.

One approach to the problem of material exploration is a quantum chemical simulation, such as a density functional theory (DFT)-based method, because many properties of materials stem from atomistic-level phenomena. However, quantum chemical calculations generally require enormous computational resources, limiting the practical use of this method in material discovery for two reasons. First, phenomena of interest in real-world applications often involve temporal and spatial scales vastly exceeding the limitations of quantum calculations, which are usually several hundreds of atoms at a sub-nanosecond scale. Second, many simulations are required to explore the configurational space during computational material discovery.

To address these challenges, several alternate computational models have been developed to directly estimate the potential energy surface of an atomic structure. For example, conventional methods called empirical potentials, which model the interaction between atoms as a combination of analytic functions, have been developed with some success, including for simple pairwise models \cite{jones1924determination}, metals \cite{PhysRevB.29.6443,doi:10.1080/01418618408244210}, covalent bonds \cite{PhysRevB.39.5566}, and reactive phenomena.\cite{doi:10.1021/jp004368u,senftle2016reaxff}
More recently, some machine learning-based approaches have been proposed, including Gaussian processes \cite{PhysRevX.8.041048,PhysRevLett.104.136403,PhysRevLett.108.058301} and support vector machines.\cite{10.1109/INCoS.2013.26}

In recent years, neural network potentials (NNPs) have rapidly gained attention owing to the high expressive power of neural networks (NNs) combined with the availability of large-scale datasets. As datasets and models evolve, the scope of NNP applications has gradually expanded. As a benchmark for molecular systems, the QM9 dataset\cite{qm9v1,qm9v2}, which covers possible patterns of small molecules, has been widely used. Initially, NNPs for organic molecules have focused on H, C, N, and O, which are the major elements in organic molecules. In subsequent studies, NNPs have been extended to include elements such as S, F, and Cl.\cite{ani1,ani2x} For NNPs targeting crystal structures \cite{cgcnn,megnet}, the Materials Project\cite{MaterialsProject}, a large-scale materials database based on DFT calculations, is often used as a benchmark dataset.
The Open Catalyst Project, which targets molecular adsorption in catalytic reactions, has constructed a massive surface adsorption structure dataset known as the Open Catalyst 2020 (OC20) dataset.\cite{oc20intro,oc20data}
In this way, the area covered by NNPs has gradually expanded.

However, significant challenges remain in the application of NNPs to computational material discovery. One unsolved issue is how to achieve the generalization needed to accurately assess the properties of unknown structures.
All previously proposed datasets were generated based on known structures, and thus models trained using such datasets are only applicable to a limited configurational space.
For example, the Open Catalyst Project have clearly stated that previous datasets are inappropriate for their adsorption task.
By defining the system to be simulated in advance, the local configuration of atoms and combinations of elements to be generated can be reduced, thus significantly decreasing the difficulty in creating the model. However, as a disadvantage of this approach, it is necessary to recreate the NNPs and datasets for each structure of interest.

In contrast to the tasks described in previous datasets, simulations of unknown or hypothetical materials are quite common in the process of material exploration.
Thus, limiting the target domain to existing materials is undesirable.
This is where a major gap exists between the requirements for current NNPs and material exploration.
This gap is analogous to the difference between specific object recognition and general object recognition in computer vision.

It was recently demonstrated that the NN losses in various tasks follow a power law well based on the size of the dataset and the number of NN parameters when applying a suitable model, regardless of the target domain.\cite{kaplan2020scaling,liu2021pay}
Thus, NNs can achieve a high accuracy even with datasets having high diversity.
This result indicated that there is a way to overcome this challenging task through the use of a sufficient dataset and architecture.

We applied the above concept to the development of an NNP.
Instead of collecting realistic, known stable structures, we aggressively gathered a dataset containing unstable structures to improve the robustness and generalization ability of the model. The dataset includes structures with irregular substitutions of elements in a variety of crystal systems and molecular structures, disordered structures in which a variety of different elements exist simultaneously, and structures in which the temperature and density are varied.
The NNP architecture was also designed under the premise of this highly diverse dataset.
The architecture should treat many elements without a combinatorial explosion.
In addition, it can utilize higher-order geometric features and handle the necessary invariances.

In this study, we created a universal NNP, called PreFerred Potential (PFP), which is capable of handling any combination of 45 elements selected from the periodic table. We conducted simulations using PFP for a variety of systems, including i) lithium diffusion in LiFeSO${}_4$F, ii) molecular adsorption in metal-organic frameworks, iii) a Cu-Au alloy order--disorder transition, and iv) material discovery for a Fischer--Tropsch catalyst. All results demonstrated that PFP produces a quantitatively excellent performance.
All results were reproduced using a single model in which no prior information regarding these four types of systems was applied as a prerequisite for training.

\section{Results}

\subsection{Lithium diffusion}
The first example application is lithium diffusion in lithium-ion batteries.
Lithium-ion batteries are used in various applications, such as portable electronic devices and electric vehicles.
The demand for lithium-ion batteries has been increasing in recent decades, and new battery materials have been explored.
One of the essential properties of lithium-ion batteries is their charge-discharge rate.
Faster lithium diffusion, that is, a lower activation energy of lithium diffusion, leads to faster charge and discharge rates.
DFT calculations have been widely applied to lithium-ion battery materials \cite{van1998first,yamada2016hydrate}, and the activation energies of lithium diffusion have also been calculated for various materials.\cite{morgan2003li,he2017origin}
An activation energy calculation requires accurate transition state estimations, as well as the initial and final states.
The transition state is a first-order saddle point in the reaction pathway between the initial and final states.
To correctly obtain the structure and energy of the transition state, a smooth and reproducible potential is required, even near the first-order saddle point, which is far from the geometrically optimized structures and harmonic vibration.
The nudged elastic band (NEB) method \cite{JONSSON1998} is one of the most widely used methods for obtaining the reaction path, and an improved version of this method, climbing-image NEB (CI-NEB) \cite{henkelman2000climbing}, can be used to obtain the transition state.

The tavorite-structured $\mathrm{LiFeSO_4F}$ ($P\overline{1}$) is a cathode material for lithium-ion batteries with a high voltage of 3.6 V.\cite{recham20103}
According to existing DFT calculations, this material shows a one-dimensional diffusion, that is, the low activation energy of lithium diffusion in only a single direction.\cite{mueller2011evaluation}
We calculated the activation energy of lithium diffusion in $\mathrm{LiFeSO_4F}$ using the CI-NEB method using PFP and compared the results with those of the existing DFT calculations. It is noted that neither the crystal structure of $\mathrm{LiFeSO_4F}$ nor that of $\mathrm{FeSO_4F}$ are included in the dataset.

A delithiated structure of $\mathrm{LiFeSO_4F}$, that is, the structure of $\mathrm{FeSO_4F}$, is obtained by removing all lithium in the $\mathrm{LiFeSO_4F}$ unit cell and then geometrically optimizing the cell parameters and site positions while maintaining the symmetry.
All CI-NEB calculations were conducted with one lithium atom and a $2\times2\times2$ supercell of $\mathrm{FeSO_4F}$.
The chemical formula is $\mathrm{Li_{1/16}FeSO_4F}$.
The cell parameters are frozen to those of $\mathrm{FeSO_4F}$.
The diffusion paths in the [111] and [101] directions contain three diffusion hops for each, and the diffusion path in the [100] direction contains one diffusion hop.\cite{recham20103}
There are nine NEB images for each CI-NEB calculation.
PFP conducts all of this calculation on a single GPU in approximately 5 min.

In addition, MD simulations were performed to confirm the results of the CI-NEB calculation and demonstrate that PFP can be used for the finite-temperature dynamics simulation. The same structure as the initial state of the CI-NEB calculation was used for MD simulations. The temperature was set at 300 K, 325 K, 350 K, 375 K, and 400 K. Eight trajectories of 100 ps were generated for each temperature. The details of the MD simulation settings and the calculation method for the activation energy are described in Supplementary Data 13.

\begin{figure}[tbp]
    \centering
    \includegraphics[width=0.99\linewidth]{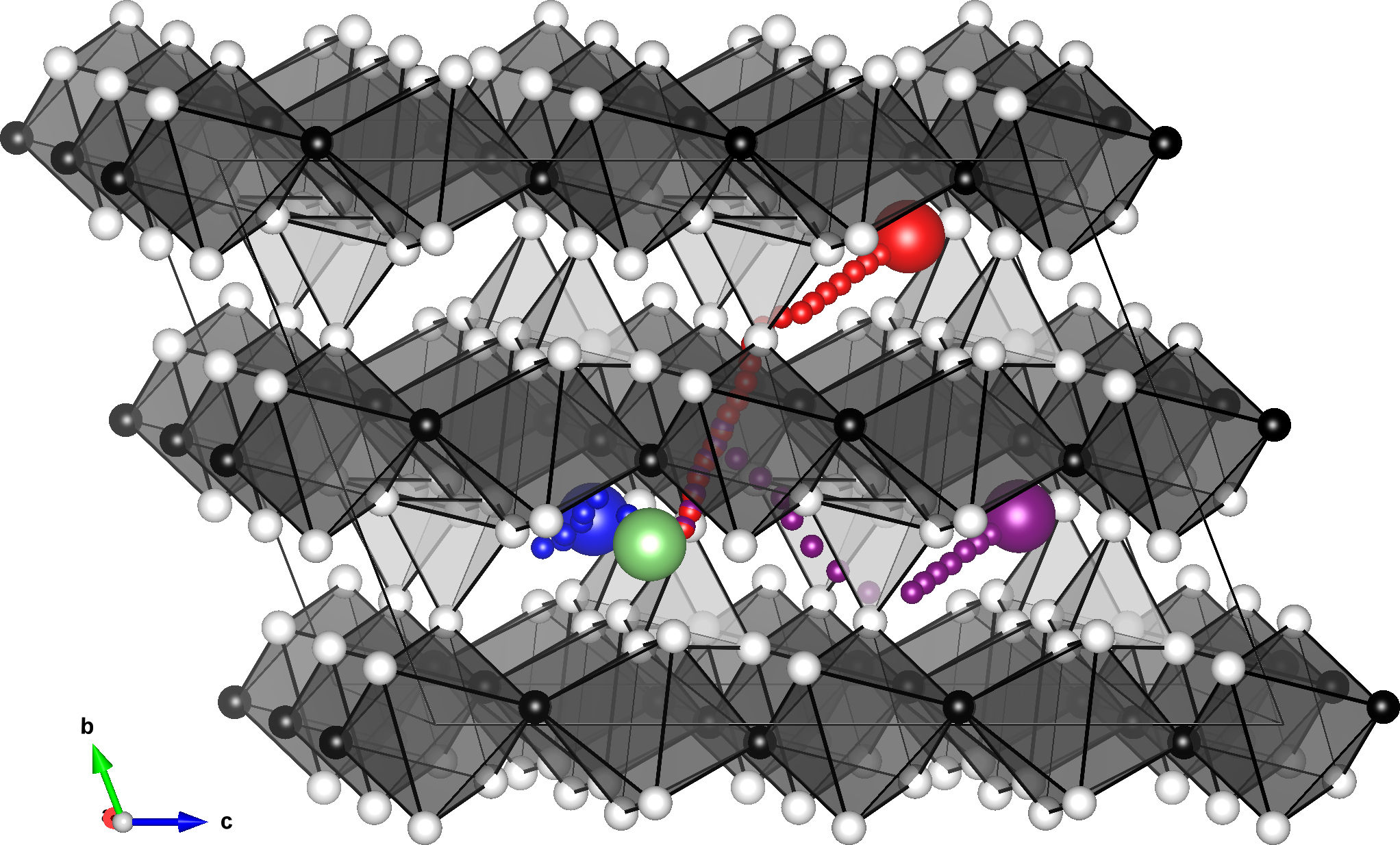}
    \caption{
    Lithium diffusion paths projected onto a $2\times2\times2$ supercell of $\mathrm{FeSO_4F}$.
    Elements are represented by white spheres (oxygen), black spheres (fluorine), dark gray octahedra (iron), and light gray tetrahedra (sulfur). The small red spheres represent the lithium diffusion path in the $[111]$ direction, from the large green sphere (initial lithium site) to the large red sphere (final lithium site).
    The diffusion paths in the $[101]$ and $[100]$ directions are represented by purple and blue spheres, respectively.
    The figure is drawn using the VESTA visualization package.\cite{momma2011vesta}}
    \label{fig:Li_Diffusion}
\end{figure}

The obtained lithium diffusion paths are shown in Fig. \ref{fig:Li_Diffusion}, and the activation energies are shown in Table \ref{tab:Li_Diffusion}.
The PFP qualitatively reproduces a DFT result in which $\mathrm{LiFeSO_4F}$ exhibits one-dimensional diffusion.
Furthermore, quantitatively, the PFP reproduces the DFT result with high accuracy.
Although neither transition states nor reaction pathways are explicitly given in the training data for creating PFP, it is possible to correctly infer the energies of the transition states far from a stable state, as well as harmonic oscillations from such state.

\begin{table}[tbp]
    \centering
    \begin{tabular}{cccc}
    & \multicolumn{3}{c}{activation energy (eV)} \\
    method  & $[111]$ & $[101]$ & $[100]$ \\ \hline
    DFT \cite{mueller2011evaluation} & 0.208 & 0.700 & 0.976 \\
    PFP (NEB) & 0.214 & 0.677 & 1.015 \\
    PFP (MD) & 0.202 & - & - \\
    \end{tabular}
    \caption{
    Activation energies for lithium diffusion through $\mathrm{LiFeSO_4F}$ at the dilution limit (i.e., through $\mathrm{FeSO_4F}$). 
    Note that DFT values are calculated without Hubbard U corrections,\cite{dudarev1998electron} although our datasets were calculated based on the corrections.
    The tests conducted by Muller et al. indicate that the corrections do not significantly affect the predicted activation energies. \cite{mueller2011evaluation}}
    \label{tab:Li_Diffusion}
\end{table}

\subsection{Molecular adsorption in metal-organic framework}

Metal-organic frameworks (MOFs) are a class of nanoporous crystalline materials with exceptionally high surface area. They consist of metal centers bridged by organic linkers, thereby creating diverse crystalline structures with a wide range of elements. Thus, these materials are ideal for testing the capability of PFP owing to their complex chemical structures containing organic and inorganic parts with unique crystalline pore structures. Such a system is normally difficult to reproduce using a conventional classical interatomic potential without finetuning the potential parameters. Quantum chemical calculations, such as the DFT approach, may avoid such issues in exchange for tremendous computational costs.

To test the applicability of PFP to MOFs, some representative materials were selected, and the cell geometries were optimized. Here, it should be emphasized that none of the MOF structures are included in our training dataset; thus, this is an out-of-domain test of our model. The starting crystalline structures were obtained from the Cambridge Structure Database (CSD).\cite{Groom:bm5086} The initial structures were cleaned by removing the physically adsorbed molecules in the pores of the MOFs. Water molecules that are chemically bound to the metal centers are maintained. These structures are referred to as ``hydrated'' structures. Other minor cleansing procedures were performed by adding hydrogen atoms to the framework and removing overlapping atoms to ensure physically reasonable crystal structures and stoichiometries. Dispersion interactions were also considered. The Grimmes D3 model was adopted for this purpose\cite{Practical_Methods_of_Optimization}. Notably, the dispersion correction can be calculated separately from the DFT, and adding it to PFP is still effective from a view of calculation time. To maximize the efficiency of the dispersion correction calculation, we implemented the GPU-accelerated version of DFT-D3 using PyTorch \cite{pytorch} and made it open-source and freely available (https://github.com/pfnet-research/torch-dftd).
Details of the calculation setup are provided in Supplementary Data 14.

\begin{figure}[tbp]
\centering
\includegraphics[width=0.9\linewidth]{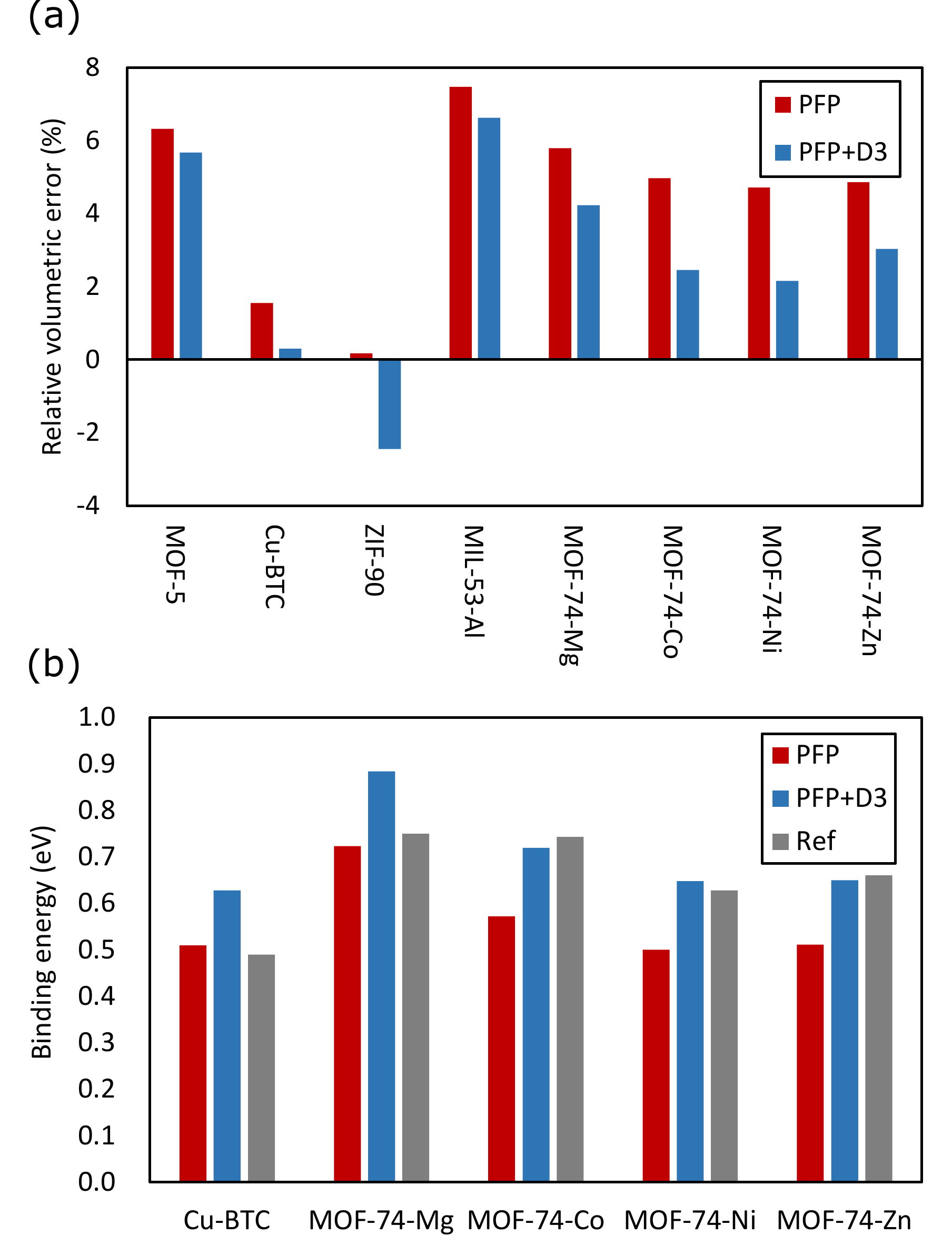}
\caption{(a) Relative error between optimized unit cell and experimentally determined cell volumes. (b) Mean binding energies of H2O molecules in selected MOFs with open metal sites with PFP, PFP+D3, and reference values. All reference values are obtained from DFT calculations.}
\label{fig:mof_all}
\end{figure}

The PFP-optimized crystal structures were compared with the experimental crystalline structures reported in the literature. Figure \ref{fig:mof_all} (a) shows the relative error in the cell volume of the MOF crystals. The individual cell parameters are provided in Supplementary Data 15. The predicted and experimental lattice parameters are in good agreement, and the mean absolute error of the cell volume is +4.5 \% and + 3.4 \% with and without dispersion corrections, respectively. This translates to a deviation in the lattice parameters of approximately +0.7 \% for both cases. The results are encouraging because a good agreement is obtained, although MOFs are out-of-domain datasets, and no such structure is used to train the PFP.

Some MOFs have unsaturated open-metal sites that are active for the chemisorption of small molecules. For example, MOF-74 is a MOF with a one-dimensional pore structure consisting of metal(M)-oxide nodes bridged by a DOBDC ligand (DOBDC = 2,5-dioxido-1,4-benzenedicarboxylate) \cite{https://doi.org/10.1002/chem.200701370}.
It is one of the early generations of MOFs, and its unique structure and properties have been well-studied\cite{Furukawa1230444}. There are different versions of MOF-74 with Ni, Co, Mg, and Zn, as well as of their combinations as the metals.
The metal node is normally coordinated with water molecules because of the hydrothermal synthesis. The sample needs to be dehydrated by annealing at 200 \textdegree{}C to remove the water molecules and create open metal sites. These sites can be the locations for the adsorption of various small molecules and may act as metal centers for catalytic reactions. Another well-known example of MOFs with open metal sites is Cu-BTC (Cu${}_3$(BTC)${}_2$, where BTC = benzene-1,3,5-tricarboxylate)\cite{doi:10.1126/science.283.5405.1148}. Cu-BTC contains a copper-oxide node linked by BTC. These copper nodes can be activated by removing the chemisorbed molecules. These systems are a good test ground for the fidelity of PFP for molecular adsorption in nanoporous materials.

The mean binding energy of a water molecule is given by:

\begin{equation}
    \begin{aligned}
        \Delta E=&-E\left(\mathrm{MOF}+N_{\mathrm{H_2O}}\times\mathrm{H_2O}\right)/N_{\mathrm{H_2O}} \\
        &+E\left(\mathrm{MOF}\right)/N_{\mathrm{H_2O}}+E\left(\mathrm{H_2O}\right),
    \end{aligned}
\end{equation}

where $E\left(\mathrm{MOF}+N_{\mathrm{H_2O}}\times\mathrm{H_2O}\right)$, $E\left(\mathrm{MOF}\right)$, and $E\left(\mathrm{H_2O}\right)$ are the total energies of the fully hydrated, dehydrated, and isolated water molecules, respectively. In addition, $N_{\mathrm{H_2O}}$ is the number of water molecules in the system, which is 18 for all cases. Based on this definition, the more stable the compound, the more positive $\Delta E$.

Figure \ref{fig:mof_all} (b) displays the mean binding energies of water molecules in the selected MOFs with open-metal centers. The agreement between our predictions and those found in the literature is quite impressive. The largest deviation is in the case of Mg, where the error is more than 10 \%, whereas all other cases remain within a few percent points on average. For MOF-74 series, the agreement is better with PFP+D3. This is consistent with the fact that the literature reports use vdw-DF as the DFT functional. Conversely, in the case of Cu-BTC, the result is nearly identical to that of PFP. However, this reference uses PBE functional only, and there is no dispersion correction applied. Therefore, this is also consistent with our observation. Most importantly, PFP correctly predicts the trend in the binding energy of water molecules in a quantitative fashion. 

It should be emphasized that neither the MOFs nor the metal-organic complexes examined
in this section are explicitly provided in the training dataset for creating the PFP. Therefore, PFP learned to correctly predict the interaction between the metal centers and water molecules in such structures from the energies and forces of isolated molecules and periodic solids.

\subsection{Cu-Au alloy order-disorder transition}
Some precious metal alloys are well known for their catalytic activity, and extensive experimental and theoretical studies have been conducted. For example, gold-copper alloys are well-studied catalysts for the oxidation of CO and selected alcohol. \cite{LIU2011103,LI2012146,doi:10.1021/cs501923x}

Local microscopic structures and atomic arrangements are essential for the performance of the catalyst. The Cu-Au alloy is a particularly interesting example because it is fully miscible over a wide composition range and exhibits an order-disorder transition.\cite{doi:10.1021/nl503584q} The critical temperature is known to depend on the composition of the alloy and has been well-studied in the literature.\cite{C7NR00028F}

To demonstrate the applicability of PFP, we conducted Metropolis Monte Carlo (MC) simulations to investigate the transition temperature between ordered and disordered phases at various compositions of Cu-Au alloy. The calculations were applied at three different compositions: CuAu${}_3$, CuAu, and Cu${}_3$Au for their well-defined ordered structures. Each unit cell was expanded to $4\times 4\times 4$ unit cells and used as the starting geometry. The details of MC moves are shown in Supplementary Data 16.

\begin{figure}[tbp]
\centering
\includegraphics[width=0.73\linewidth]{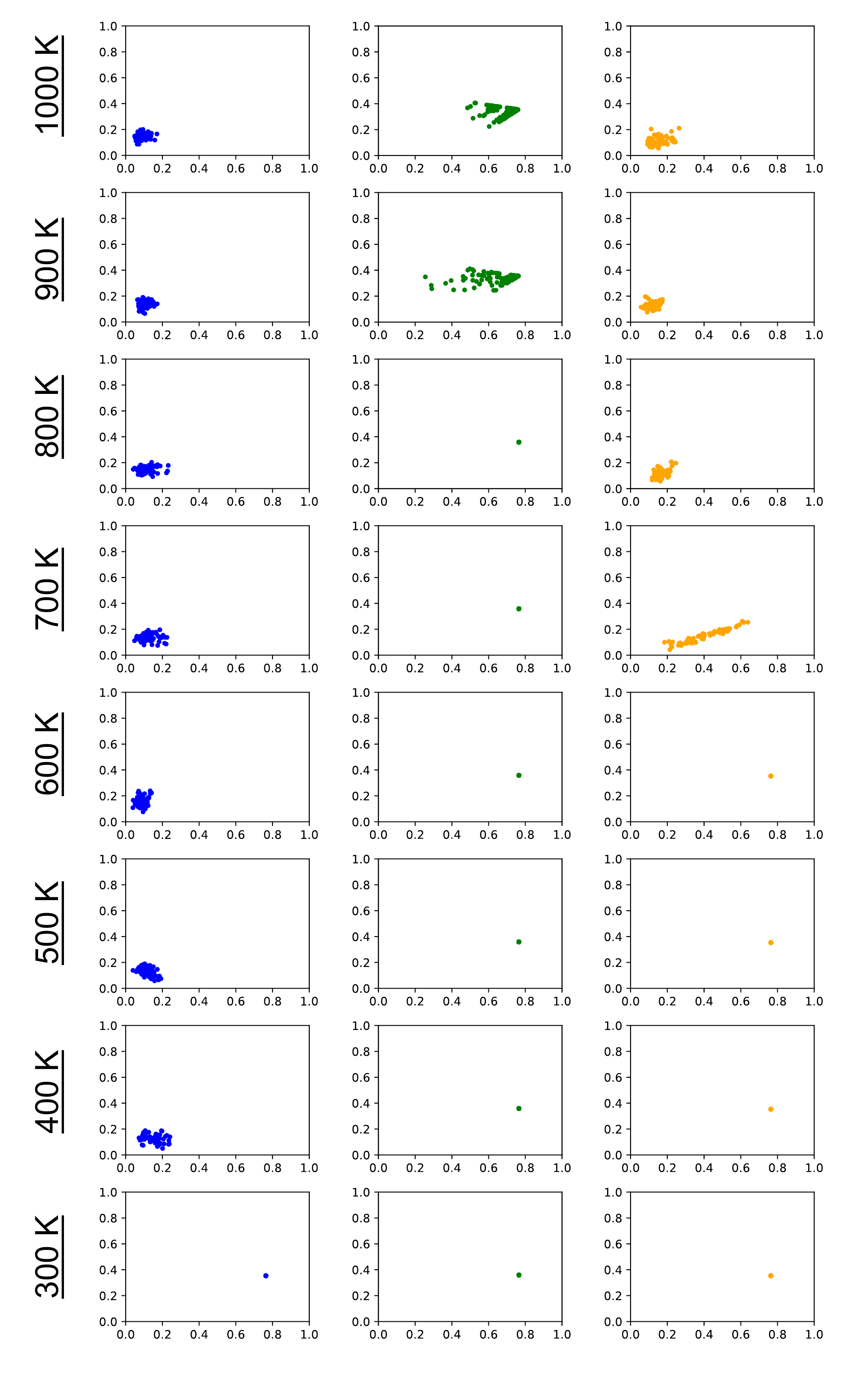}
\caption{Voronoi weighted Steinhardt parameters of CuAu${}_3$ (left), CuAu (center), and Cu${}_3$Au (right). The ordinate and abscissa of each plot represent q4 and q6, respectively. These order parameters are calculated with respect to Cu in the case of CuAu3 and CuAu, and with respect to Au in the case of Cu3Au. The disordered structures can be observed as the diffused points in the figures.}
\label{fig:cuau_2}
\end{figure}

The characterization of the resulting structures from MC simulations is summarized in Figure \ref{fig:cuau_2}.
The computed order parameters show a clear transition from ordered to disordered phases. Perfectly ordered structures at low temperatures have well-defined order parameters and can be seen as a single point. By contrast, as the temperature increases, disturbances appear, and the plot becomes dispersed. The calculated transition temperatures are 300--400 K for CuAu${}_3$, 800--900 K for CuAu, and 600--700 K for Cu${}_3$Au. These trends are consistent with the reported transition temperatures (CuAu${}_3$, 440--480 K; CuAu, 670--700 K; Cu${}_3$Au, 660--670 K\cite{C7NR00028F}) and demonstrate the applicability of PFP.

\subsection{Material discovery for a Fischer--Tropsch catalyst}
Another example of the power of PFP is given in the context of a heterogeneous catalysis. The Fischer--Tropsch (FT) reaction is a synthesis of hydrocarbons from hydrogen and carbon monoxide, involving a wide variety of elementary chemical reactions.\cite{DRY2002227,doi:10.1021/acscatal.0c02420} This reaction process is particularly important for the generation of fuel from renewable and sustainable energy sources. In this example, we focus our attention on the methanation reactions and CO dissociation processes on Co surfaces.

The methanation reactions of synthesis gases are well documented in the literature.\cite{ZIJLSTRA2020131} In particular, 20 elementary reactions on the Co($0 0 0 1$) surface have been examined, and corresponding activation energies are compared with the values reported in the literature.

Each simulation cell geometry consisted of 45 Co atoms with 5 atomic layers. Only the bottom three layers were constrained, and the rest were allowed to relax. The vacuum size was set to 10 \angstrom \ ($1 \,\mathrm{\angstrom} = 10^{-10}\,\mathrm{m}$). The geometry is optimized until the maximum force of all atoms reaches below 0.05 eV/\angstrom. The activation energy was determined by CI-NEB using 14 images for each process. Zero-point energy corrections were also included in the calculations.

\begin{figure}[tbp]
\centering
\includegraphics[width=0.95\linewidth]{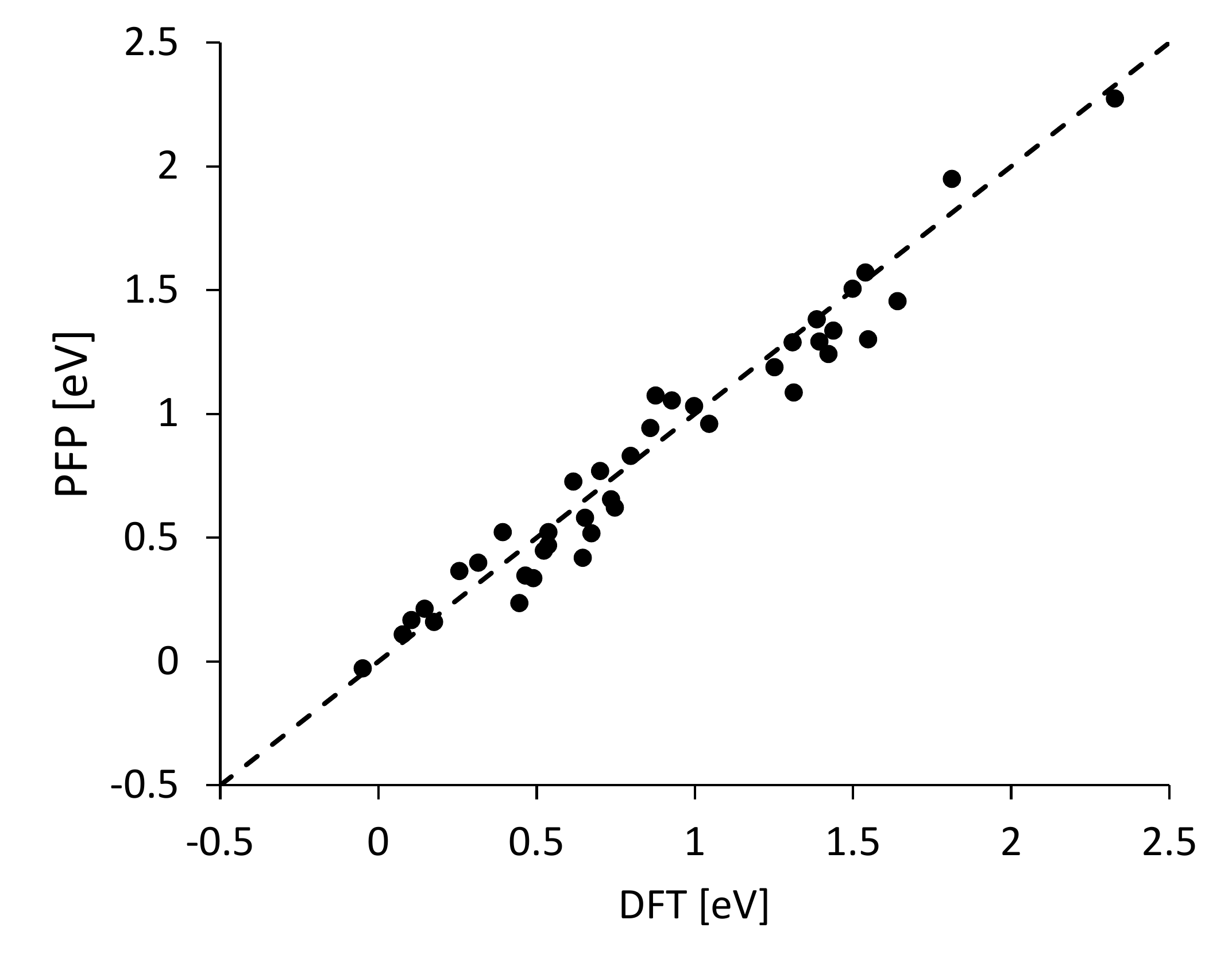}
\caption{Comparison of the activation energies of methanation reactions of synthesis gas on Co($0 0 0 1$). The ordinate and abscissa represent the PFP prediction and reference DFT values, respectively. The zero-point energy corrections of the transition states are also included in the data.}
\label{fig:ft_1}
\end{figure}

Figure \ref{fig:ft_1} shows a comparison of the computed activation energies between PFP and the reported values.\cite{ZIJLSTRA2020131} The correlation coefficient is 0.98, and the mean absolute error is 0.097 eV, indicating the high fidelity of PFP for the prediction of activation energies in this class of chemical reactions.

Backed with the high fidelity of PFP, we explored possible promoter elements for the CO dissociation reaction on a Co surface. CO dissociation is a critical part of the overall reaction mechanism of the FT process. Although it was reported to be approximately 1 eV for the activation energy of pure Co surfaces, a reduction of the activation barrier is desired, and several efforts have been reported in the literature.\cite{doi:10.1021/acscatal.9b01967} However, DFT calculations for such exploration demand a high computational cost, and PFP can accelerate such a screening process. Specifically, we explored the CO dissociation reaction pathways by CI-NEB on the Co($1 1 \overline{2} 1$) step surface. In the promoter search process, a Co atom was randomly replaced with a promoter element, and the CI-NEB calculations were repeated over the surface. The CI-NEB was repeated 10 times on each surface, and a list of activation energies was obtained.

\begin{figure}[tbp]
\centering
\includegraphics[width=0.98\linewidth]{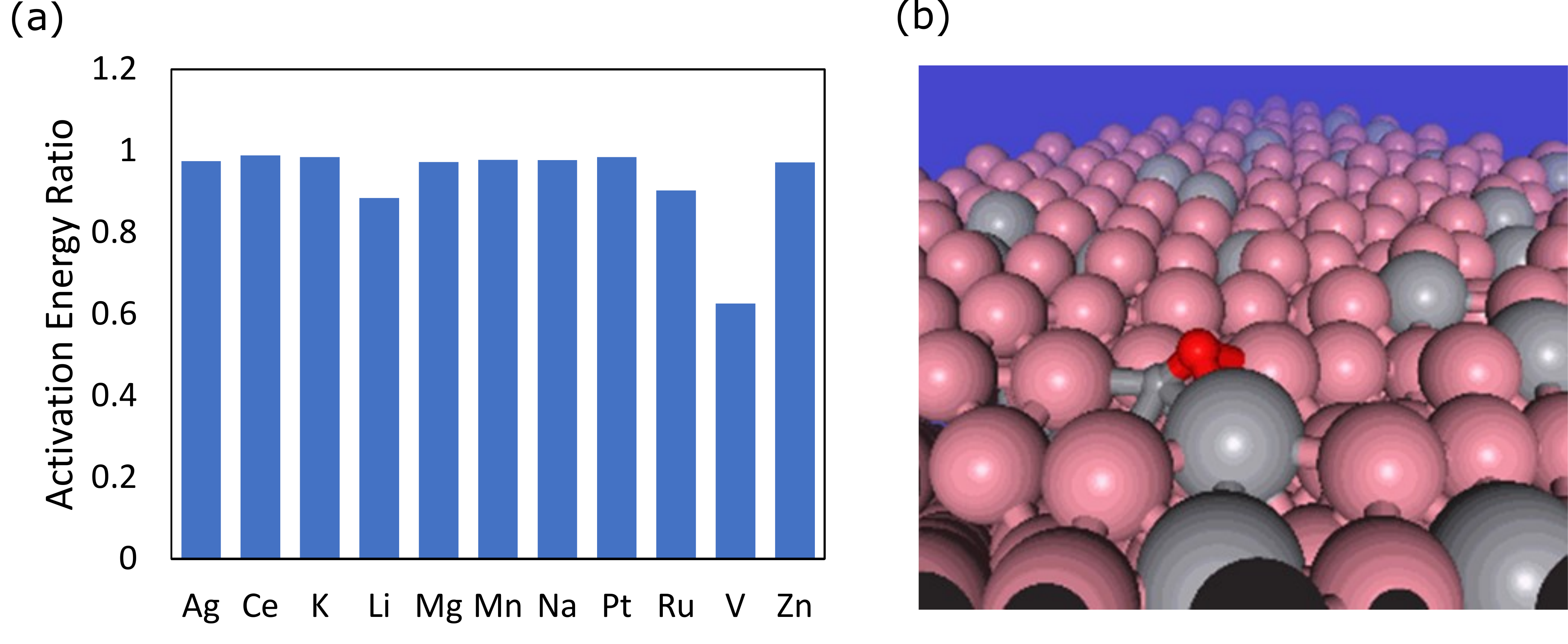}
\caption{(a) Normalized activation energies of CO dissociation. (b) CO adsorbed configuration of a Co($1 1 \overline{2} 1$) surface with V promoters. The representative atoms are Co (pink), O (red), C (small gray), and V (big gray).}
\label{fig:ft_2}
\end{figure}

Because they are often found in the literature as promoters of certain reactions, we chose the following 11 elements (Ag, Ce, K, Li, Mg, Mn, Na, Pt, Ru, V, and Zn) for our study. The results are summarized in Figure \ref{fig:ft_2}(a). Among the list, the most significant reduction (approximately 40\%) was found with V, whereas the others showed a minor impact on the activation barrier. The lowest energy configuration of CO adsorbed Co($1 1 \overline{2} 1$) with V is shown in Figure \ref{fig:ft_2}(b). The CO molecule was found to lie across the Co and V bridge sites. In fact, some experimental studies have already reported a significant reduction in the activation energy of Co by V, although we identified the element without any prior knowledge from the literature.\cite{wang2006effect,SHIMURA20151} The agreement between our findings and the literature is consistent. It is encouraging to note that our approach can facilitate the use of PFP in complex systems such as a heterogeneous catalysis.

\section{Discussion}
We developed a universal NNP called PFP, which operates on systems with any combination of 45 elements.

The results indicate that a single NNP model can describe a diverse set of phenomena with high quantitative accuracy and low computational cost.
In addition, it was also shown that PFP can reproduce structures and energetic properties that were not envisioned during the design phase. The detailed correspondence between the results and the PFP dataset is shown in Supplementary Data 11.
Our results suggest that the approach to constructing a unified NNP, instead of training an independent NNP for each target task, is promising.
Further comparison of calculation time between PFP and DFT is included in Supplementary Data 6. Although DFT calculations or other electronic structure calculations from first principles are still considered to be reliable because of the strong physics background, PFP can greatly mitigate another limitation of atomistic simulations caused by the time and space scales. The combined study of DFT and PFP or experiments using PFP-based screening will also accelerate the field of material discovery.

The result of the Fischer--Tropsch catalyst is an example of applying PFP to an actual material discovery task.
This is a typical case in which NNP is able to achieve the following three properties at the same time: (1) the ability to handle a wide variety of elements, (2) the ability to handle phenomena that were not assumed at the time of training, and (3) a significantly faster speed than that of DFT.

These results further confirm that PFP is versatile and applicable for screening a wide range of materials without prior knowledge of the atomic structures in the target domain.

\section{Method}

\subsection{Dataset} \label{sec:dataset}

\subsubsection{Systems and Structures}

In this study, we generated an original dataset which covers various systems. See Supplementary Data 10 for the definition of each subcomponents and the detailed calculation conditions on how to generate them, and Supplementary Data 19 for the statistical information of our dataset. The summary of our dataset is shown below.

Early examples of large datasets with quantum chemical calculations include QM9 \cite{qm9v1,qm9v2} and the Materials Project.\cite{MaterialsProject}
They were generated by conducting DFT calculations on various molecules or inorganic materials and collecting physical properties in geometrically optimized structures to accelerate drug or material discovery.
Although they have been utilized for predicting physical properties such as HOMO-LUMO gaps or formation energies of optimized structures, they are insufficient for generating universal potentials for new material discovery because they mainly focus on optimized structures.
In particular, the reaction, diffusion, and phase transitions are dominated by structures far from the optimized structures.
By contrast, it is unsuitable to sample geometrically random structures.
Because the probability distribution of the structures follows a Boltzmann distribution, geometrically random structures that tend to show much higher energies compared to optimized structures rarely appear in reality.
Therefore, it is important to cover as many diverse structures as possible while limiting those showing valid energies.  

To achieve this, ANI-1 \cite{ani1}, ANI-2x \cite{ani2x}, and tensor-mol \cite{tensormol} sampled not only geometrically optimized structures of various molecules, but also their surrounding regions using NMS, MD, or meta-dynamics.
Using these methods, we can obtain datasets to generate the potential to reproduce phenomena with large structural deformations, such as protein-drug docking, which is important in drug discovery.
However, these datasets focus only on molecules and do not cover systems such as crystals and surfaces.
One recent study that deserves attention is OC20 \cite{oc20data}, which has an order of magnitude larger number of data than previous studies.
Nevertheless, this dataset also focuses on catalytic reactions and only contains data on the adsorbed structures.
As we have shown, it is worth noting that these adsorbed structures are generated with known stable structures. As a result, the accuracy of the energy predictions is much lower for structures that depart from known stable structures.

Following these insights and issues, we generate an original dataset that covers all systems with molecular, crystal, slab, cluster, adsorption, and disordered structures, as shown in Table \ref{tab:datasets}.
For each system, we sampled various structures, such as geometrically optimized structures, vibration structures, and MD snapshots, to collect the data necessary to obtain a universal potential.

Our dataset consists of a molecular dataset calculated without periodic boundary conditions, and a crystal dataset calculated with periodic boundary conditions.
Each dataset contains the structure and corresponding total energies and forces obtained through DFT calculations.
The crystal dataset also includes the atomic charges.
The molecular dataset supports nine elements: H, C, N, O, P, S, F, Cl, and Br.
There is maximum of eight atoms from among C, N, O, P, and S in a molecule.
In addition to stable molecules, unstable molecules and radicals are also included.
Various structures are generated for a single molecule through geometrical optimization, NMS, and MD at high temperatures.
The two-body potentials for almost all combinations of up to H -- Kr are also calculated as additional data.
For the crystal dataset, 45 elements are supported, as shown in Fig. \ref{fig:pp}.
This includes a variety of systems, such as bulk, cluster, slab (surface), and adsorption on slabs.
Non-stable structures, such as Si with simple cubic ($Pm3m$) or FCC ($Fm3m$) structures or NaCl with a zincblende structure ($F\overline{4}3m$), as well as non-optimized structures, are also included in the crystal dataset.
For the bulk, cluster, and slab, we generated structures by changing the cell volumes or shapes, or by randomly displacing the atomic positions, instead of applying the NMS method.
For the adsorbed systems, we generated structures with randomly placed molecules in addition to the structure-optimized ones using PFP.
Disordered structures are generated using MD at high temperatures for randomly selected and placed atoms.
Molecules are also included in the crystal dataset.
The two-body potentials for almost all combinations of up to H--Bi were also calculated.
The computational resources used to acquire these datasets were approximately $6\times 10^4$ GPU days.

We provide an atomic structure dataset called the high-temperature multi-element 2021 (HME21) dataset, which consists of a portion of the PFP dataset (Supplementary Script 2). See Supplementary Data 17 for further details.

\begingroup
\squeezetable
\begin{table*}[tbp]
    \centering
    \begin{tabular}{c|cccccc|cccc|cc}
        & \multicolumn{6}{c|}{systems} & \multicolumn{4}{c|}{structures} & \multicolumn{2}{c}{\# of} \\
        Dataset & molecule & bulk & cluster & slab & adsorption & disorder & opt. & vib. & MD & TS & Elements & Data \\ \hline
        Materials Project~\cite{MaterialsProject} & & \checkmark  & & \checkmark & & & \checkmark & & & & - & $>1\times10^5$ \\
        OQMD~\cite{oqmd} & & \checkmark  & & & & & \checkmark & & & & - & $8\times10^5$ \\ 
        NOMAD~\cite{nomad} & & \checkmark  & & & & & \checkmark & & & & - & $>5\times10^7$ \\
        Jarvis-DFT~\cite{jarvis} & & \checkmark  & & & & & \checkmark & & & & - & $>4\times10^5$ \\
        AFLOW~\cite{aflow} & & \checkmark  & \checkmark & & & & \checkmark & & & & - & $>3\times10^6$(*1) \\ 
        QM9~\cite{qm9v1,qm9v2} & \checkmark & & & & & & \checkmark & & &  & 5 & $1\times10^5$ \\ 
        PubChemQC~\cite{pubchemqc} & \checkmark & & & & & & \checkmark & & &   & 30 & $>3\times10^6$  (*2) \\ 
        MD17~\cite{md17} & \checkmark & & & & & & & \checkmark & & & 4 & $9\times10^6$ \\ 
        $S_N2$ reactions~\cite{physnet} & \checkmark & & & & & & \checkmark & & \checkmark & \checkmark & 6 & $4\times10^5$ \\ 
        ANI-1~\cite{ani1} & \checkmark & & & & & & \checkmark & \checkmark & \checkmark & & 5 & $2\times10^7$ \\ 
        ANI-2x~\cite{ani2x} & \checkmark & & & & & & \checkmark & \checkmark & \checkmark & & 7 & $9\times10^6$ \\ 
        COMP6v2~\cite{ani2x} & \checkmark & & & & & & \checkmark & \checkmark & \checkmark & & 7 & $2\times10^5$ \\ 
        tensor-mol 0.1 water~\cite{tensormol} & \checkmark & & & & & &  &  & \checkmark & & 2 & $4\times10^5$ \\ 
        tensor-mol 0.1 spider~\cite{tensormol} & \checkmark & & & & & &  &  & \checkmark & & 4 & $3\times10^6$ \\ 
        TeaNet~\cite{teanet} & \checkmark & & & & & \checkmark & & & \checkmark & & 18 & $3\times10^5$ \\ 
        OC20~\cite{oc20intro,oc20data} & & & & & \checkmark & &\checkmark & \checkmark & \checkmark &  & 56(*3) & $1\times10^8$ \\ 
        PFP molecular dataset (ours) & \checkmark &  &  &  &  &  & \checkmark & \checkmark & \checkmark & & 9 & $6\times10^6$ \\
        PFP crystal dataset (ours) & \checkmark & \checkmark & \checkmark & \checkmark & \checkmark & \checkmark & \checkmark & \checkmark & \checkmark & & 45 & $3\times10^6$ \\
        \end{tabular}
    \caption{
    Comparison of DFT calculated datasets that can be used to train the neural network potential.
    ``-'' in \# of elements means that the element is not limited.
    (*1): The number is checked on May 24, 2021.
    (*2): The number is taken from ~\cite{pubchemqc}, and is updated weekly.
    (*3): The number was checked using only the training dataset of version 1.
    }
    \label{tab:datasets}
\end{table*}
\endgroup

\subsubsection{Training with multiple datasets}
In addition to the above molecular and crystal datasets, we used the OC20 dataset as a training dataset. This means that there are multiple datasets generated by different DFT conditions that are inconsistent with each other.
Attempting to merge these datasets simply does not yield a good performance in practice. Overlapping dataset regions with different DFT conditions may have harmed the training because each data point would have resulted in inconsistent energy surfaces.

However, because these datasets are well sampled in each area of strength, it is desirable to use as much data as possible to improve the generalization. Therefore, we assigned labels corresponding to the DFT conditions during training and trained the entire dataset concurrently.
During inference, it is also possible to select which DFT condition to infer by assigning labels in the same way as during training.
This approach makes it possible to learn multiple mutually contradictory datasets with high accuracy. In addition, as the model learns the consistent properties of all datasets and the differences in each, it is expected that domains that have only been computed in one DFT condition will be transferred to the inference under other DFT conditions. The additional benchmark is shown in Supplementary Data 20.

Considering that datasets will become even larger in the future, the mechanism for the simultaneous training of datasets with different DFT conditions will become more important.

We considered the crystal dataset as the most basic one. All applications shown in this study are calculated in the corresponding calculation mode.

\subsubsection{DFT calculation conditions}
DFT calculations for the molecular dataset are carried out using the $\omega$B97X-D exchange-correlation functional \cite{chai2008long} and the 6-31G(d) basis set \cite{ditchfield1971self} implemented in Gaussian 16.\cite{g16}
To reproduce the symmetry-breaking phenomena of the wavefunction, such as a hydrogen dissociation, we carry out unrestricted DFT calculations with a symmetry-broken initial guess for the wavefunction.
However, for geometrical optimization calculations, we carry out restricted DFT calculations.
We only consider singlet or doublet spin configurations except for diatomic potentials.

\begin{figure*}[tbp]
\centering
\includegraphics[width=0.80\linewidth]{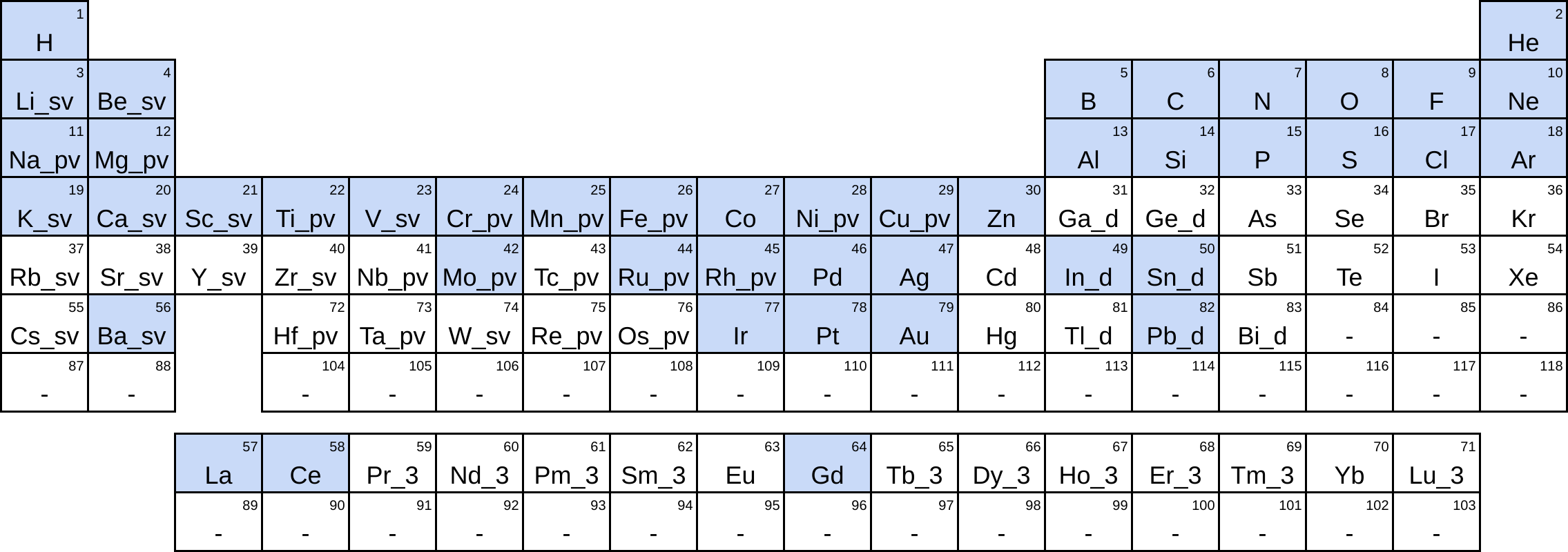}
\caption{The 45 elements supported by PFP are colored in the periodic table.
Pseudopotentials used in the DFT calculations for the PFP crystal dataset are also shown in the periodic table.
These are supplied with the VASP package, version 5.4.4, and chosen by the Materials Project \cite{MaterialsProject}.
}
\label{fig:pp}
\end{figure*}

Spin-polarized DFT calculations for the crystal dataset are carried out using the Perdew--Burke--Ernzerhof (PBE) exchange-correlation functional \cite{perdew1996generalized} implemented in the Vienna Ab-initio Simulation Package \cite{kresse1993ab,kresse1994ab,kresse1996efficiency,kresse1996efficient} (VASP), version 5.4.4, with GPU acceleration.\cite{hacene2012accelerating,hutchinson2012vasp}
The projector-augmented wave (PAW) method \cite{blochl1994projector,kresse1999ultrasoft} and plane-wave basis are used with a kinetic energy cutoff of 520 eV and pseudopotentials, as shown in Figure \ref{fig:pp}.
Here, $k$-point meshes are constructed based on the cell parameters and the $k$-point density of 1000 $k$-points per reciprocal atom. However, $\Gamma$-point-only calculations are carried out for structures with vacuum regions in all directions, such as molecules and clusters.
For the DFT calculations on a wide variety of systems, including insulators, semiconductors, and metals, under the same conditions, we use Gaussian smearing with a smearing width of 0.05 eV.
The generalized gradient approximation with Hubbard $U$ corrections (GGA+$U$) proposed by Dudarev et al. \cite{dudarev1998electron} is used with the $U-J$ parameters shown in Table \ref{table:U-J}.
To maintain the consistency of the energies and forces in the different systems, we use the GGA+$U$ method for all structures, including metallic systems.
To consider both ferromagnetism and anti-ferromagnetism, we carry out a calculation with both parallel and anti-parallel initial magnetic moments and adopt the result with the lowest energy.
Nevertheless, for some systems, we carry out the calculations using only parallel initial magnetic moments.
Bader charge analyses \cite{tang2009grid,sanville2007improved,henkelman2006fast,yu2011accurate} are carried out to obtain atomic charges.

\subsection{Trained properties}
The energy of the system, atomic forces, and atomic charges are used for the training procedure.
Atomic charges are considered as supplementary data. Although they are neither directly used to calculate energy nor to simulate the dynamics, they are expected to have information about the local environment of the atoms.

\begin{table}[tbp]
\caption{List of $U-J$ parameters. Values except for Cu are used in the Materials Project\cite{MaterialsProject}, and the value for Cu is determined by Weng et al. \cite{wang2006oxidation}}
\label{table:U-J}
\begin{tabular}{cccccccccc}
 Elements & V & Cr & Mn & Fe & Co & Ni & Cu & Mo & W \\ \hline
 U--J (eV) & 3.25 & 3.7 & 3.9 & 5.3 & 3.32 & 6.2 & 4.0 & 4.38 & 6.2
\end{tabular}
\end{table}

\subsection{Architecture}

\subsubsection{Neural network architecture}
The TeaNet\cite{teanet} architecture was used for the base NNP architecture of the PFP. The TeaNet architecture incorporates a 2nd-order Euclidean tensor into the GNN and performs message passing of scalar, vector, and tensor values to represent higher-order geometric features while maintaining the necessary equivariances. For a detailed explanation of the TeaNet architecture, such as step-by-step operation, the method of treating invariances, schematic comparison between the other models, and the reported original performances for both learning procedure and MD applications, see the original material\cite{teanet}. The benchmark score using HME21 dataset are shown in Supplementary Data 18.

To adopt the PFP dataset, several architectural modifications were made in this study. The major modifications are shown below.

First, the Morse-style two-body potential term is introduced in addition to the TeaNet architecture. The main purpose is to reproduce the short-range repulsion effect. When the distance between two atoms becomes much closer than the stable bond distance, the nuclear repulsion force becomes dominant, and the energy increases rapidly as the distance decreases. Usually, these types of structures are not observed during the dynamics simulations. In addition, the requirement of accurate energy estimations is not considered for these high-energy structures. However, if the NNP does not learn these structures, it is difficult to reproduce the above nature when the structure accidentally contains a very close atom pair. It is possible to estimate extremely low energies for these structures. As an example of the application of PFP, such a scenario may be fatal when performing exploratory atomic system calculations, such as structure sampling using Monte Carlo methods or structure estimation using generative models. From the aspect of the training procedure, extremely large values make the training process difficult. Therefore, we trained the parameters of the Morse-style two-body potential for all possible combinations of elements independently and added them to the energy term separately. As described above, this modification is aimed to expand the practical convenience. Neither the dataset nor the applications presented in this study deal with such an energetically extreme region, and it is assumed that the introduction of the two-body potential has negligible effect.

As described in the dataset section, the dataset contains multiple DFT conditions, such as different basis functions or exchange-correlation functionals. The data points are consistent at a high accuracy level under the same computational conditions but not between different computational conditions. This difference cannot be eliminated by zero-point shifts or linear multiplications of the energy. Unifying these sub-datasets directly is considered to provide unintended virtual energy gaps. To address this problem, the DFT condition is set as an additional input label during the training. Label information is also needed during inference. This is referred to as the ``calculation mode'' of PFP. Therefore, the calculation mode has two aspects. One is to enable the training of multiple datasets that have different conditions simultaneously, and the other is to provide a feature to select those conditions for users.

The output of the TeaNet architecture is modified to output atomic charges in addition to the total energy. Charges are considered auxiliary values. Unlike the charge equilibrium method, the charges are calculated using the forward path of the GNN. The explicit Coulombic interaction term was not included. This modification has two purposes. One is to allow PFP users to use the output charges for post-processing molecular dynamics. The other is to increase the number of learned properties for the same DFT calculations.

\subsubsection{NNP characteristics}
In this section, the characteristics of PFP are summarized from the perspective of NNP architecture.

PFP, or its GNN architecture TeaNet, has invariance for $\mathrm{E}(3)$ transformations. In other words, PFP holds rotational invariance, translational invariance, and mirror-image reversal invariance. In addition, PFP is a fully local interaction model. This means that the information of the local structure cannot propagate over an infinite distance.
For example, suppose there are two molecules, A and B, that are sufficiently far apart. It is guaranteed that whatever state molecule B is in (i.e., stationary, in the middle of a chemical reaction, or artificially erased at a certain moment during the simulation), molecule A is, in principle, unaffected. The number of GNN layers is 5. The cutoff distance of the GNN layer depends on the stage of the layer; they are set to 3 \angstrom, 3 \angstrom, 4 \angstrom, 6 \angstrom, and 6 \angstrom, respectively. This was determined by considering the balance between computational cost and accuracy. This can be regarded as a special case where all cutoff distances are equal to 6 \angstrom, which is the original TeaNet architecture. Since GNN is multi-layered, the information of the atoms propagates through the network to their neighbors, and thus the distance at which one atom interacts with another is the summation of those cutoff distances, which is 22 \angstrom. The physical counterpart of this phenomenon is the long-range interactions that occur as a result of the connected electron orbitals, such as metallic bonds and interactions through $\pi$-bonds.

Those properties are beneficial for improving generalization.
Since both invariances and the local interaction model are satisfied, the spatial invariances are maintained for two spatially separated molecules independently.
Furthermore, the extensive energy properties are preserved. In other words, when a system is composed of the sum of separated subsystems, the energy is also the sum of such subsystems. In addition, when the size of the system is doubled in the direction of the periodic boundary, the energy of the system is guaranteed to double.

PFP follows TeaNet's differentiable nature up to a higher order with respect to the position of the atom. The smoothness of the energy surface is a property directly related to the stability of the calculation, both in minimization calculations, such as structural relaxation calculations and NEB methods, and in long-time dynamics calculations. Furthermore, although molecular dynamics simulations use forces corresponding to first-order derivatives of energy, they often require quantities corresponding to higher-order derivatives, such as elastic modulus calculation, or minimization based on the quasi-Newton method.

The additional benchmark of the PFP architecture using OC20 dataset is shown in Supplementary Data 4.

\section{Data availability}
Source data will be available if this manuscript is accepted. The original sentences are: [Source data are provided with this paper. Simulation script files and output data corresponding to the result section are included in Supplementary Script 1. We provide an atomic structure dataset called the high-temperature multi-element 2021 (HME21) dataset, which consists of a portion of the PFP dataset in Supplementary Script 2.]

\section{Code availability}
Code will be available if this manuscript is accepted. The original sentences are: [Code for NNP architecture benchmark using HME21 is available in Supplementary Script 3. It contains TeaNet (base model of PFP) implementation with the trained parameters.
PFP is provided in the proprietary software named (NAME anonymized during the peer review process). The code and trained parameters are not open-source, but PFP can be used to reproduce the results through software-as-a-service (URL anonymized during the peer review process).]

\section{Author contributions}
Author contributions is anonymized during the peer review process.

\section{Competing interests}
Competing interests is anonymized during the peer review process.

\bibliographystyle{unsrtnat}
\bibliography{references}

\end{document}


\beginsupplement
\title{Supplementary Information}
\maketitle

\section*{Data 1: NN architecture benchmark using OC20 dataset}
As mentioned in the Introduction, the OC20 dataset targets adsorbed structures on crystal surfaces. Numerical experiments suggest that this is a more challenging task than molecular or crystal structure data. This dataset was used to evaluate the performance of the architecture of the PFP. 

First, we demonstrate the performance of the PFP architecture on the structure of energy and forces (S2EF) task \cite{oc20data}.
We used the S2EF 2M dataset as training data, which is a sub-dataset two orders of magnitude smaller than the largest dataset provided by OC20. For evaluation, we used the validation dataset, which was not used in the training process. The values of the baseline models (SchNet and DimeNet++) correspond to the test datasets.
The results are listed in Table \ref{tab:oc20_regression}. The PFP architecture showed a good performance compared with the baseline models.

It should be emphasized that even though the OC20 dataset covers a wide range of adsorbed structures, the model trained using the OC20 dataset is insufficient for the material discovery task as defined in this study.
As a demonstration, we calculated the energies and densities of various crystal structures of silicon using the PFP architecture trained using only the OC20 dataset, and DimeNet++, which has an excellent score in the existing OC20 leaderboard.

The results showed a similar trend for both models. The first problem is the estimation of a stable structure. Both the DimeNet++ and PFP architectures trained with OC20 failed to estimate that the diamond structure was the most stable in both architectures. The most stable structure is body-centered cubic (BCC) for DimeNet++ and face-centered cubic (FCC) for the PFP architecture trained with OC20.
This inconsistency may not be a problem when simulating silicon during the diamond structure phase. However, this is a problem in the material exploration task, where the other structures are predicted to be more stable.
The second problem can be observed in systems with large deformations. When plotting the energy surface against the volume change, another stable point was often found far from the original stable bond distance. See Supplementary Data 3 for the detailed results.
These problems are thought to be due to the limited diversity of structures covered by the dataset.
The results indicate that, to build a universally applicable NNP for material discovery, we need to pay attention to the variety of structures in the dataset.

\begin{table}[htbp]
  \centering
  \begin{tabular}{lD{.}{.}{5}D{.}{.}{5}D{.}{.}{5}D{.}{.}{5}}
Model     & \multicolumn{1}{c}{ID} & \multicolumn{1}{c}{OOD Ads} & \multicolumn{1}{c}{OOD Cat} & \multicolumn{1}{c}{OOD Both} \\ \hline
          & \multicolumn{4}{c}{Energy mean absolute error (MAE) [eV] (↓)} \\
SchNet    & 0.4426 & 0.4907 & 0.5288 & 0.7161 \\
DimeNet++ & 0.4858 & 0.4702 & 0.5331 & 0.6482 \\
PFP (OC20)& \multicolumn{1}{B{.}{.}{5}}{0.2258} & \multicolumn{1}{B{.}{.}{5}}{0.2345} & \multicolumn{1}{B{.}{.}{5}}{0.4044} & \multicolumn{1}{B{.}{.}{5}}{0.4762} \\ \hline
          & \multicolumn{4}{c}{Force MAE [eV/\angstrom] (↓)} \\
SchNet    & 0.0493 & 0.0527 & 0.0508 & 0.0652 \\
DimeNet++ & 0.0443 & 0.0458 & \multicolumn{1}{B{.}{.}{5}}{0.0444} & 0.0558 \\
PFP (OC20)& \multicolumn{1}{B{.}{.}{5}}{0.0418} & \multicolumn{1}{B{.}{.}{5}}{0.0453} & 0.0455 & \multicolumn{1}{B{.}{.}{5}}{0.0534} \\ \hline
          & \multicolumn{4}{c}{Force cosine (↑)} \\
SchNet    & 0.3180 & 0.2960 & 0.2943 & 0.3001 \\
DimeNet++ & 0.3623 & 0.3470 & 0.3462 & 0.3685 \\
PFP (OC20)& \multicolumn{1}{B{.}{.}{5}}{0.4848} & \multicolumn{1}{B{.}{.}{5}}{0.4743} & \multicolumn{1}{B{.}{.}{5}}{0.4559} & \multicolumn{1}{B{.}{.}{5}}{0.4888} \\ \hline
          & \multicolumn{4}{c}{EFwT (↑)} \\
SchNet    & \multicolumn{1}{B{.}{.}{5}}{0.11\%} & \multicolumn{1}{B{.}{.}{5}}{0.06\%} & \multicolumn{1}{B{.}{.}{5}}{0.07\%} & \multicolumn{1}{B{.}{.}{5}}{0.01\%} \\
DimeNet++ & 0.10\% & 0.03\% & 0.05\% & \multicolumn{1}{B{.}{.}{5}}{0.01\%} \\
PFP (OC20)& 0.02\% & 0.00\% & 0.00\% & 0.00\% \\
  \end{tabular}
  \caption{Open Catalyst 2020 S2EF task. ID and OOD refer to the ``in-domain'' and ``out-of-domain'' datasets, respectively. SchNet and DimeNet++ were extracted from the leaderboard. For the PFP, the validation dataset was used instead. See the original reference \cite{oc20data} for the definition of the tasks.}
  \label{tab:oc20_regression}
\end{table}

\clearpage
\section*{Data 2: PFP regression benchmarks for our dataset}
For the regression benchmark, we extracted three types of components from PFP dataset. See Supplementary Data 10 for the details of PFP dataset creation method.

The first type consists of disordered structures. The structure generation process is as follows: First, atoms are randomly selected from the periodic table and placed in the simulation cell. Next, the system is melted at approximately 10000 K through an MD simulation. An additional MD simulation is then conducted at 2000 K. See Supplementary Data 10 (disordered section) for details. The typical number of atoms in a single structure is 32. The structures created in this manner are expected to cover a vast range of phase spaces with little dependence on prior knowledge. This dataset is expected to provide a highly stringent assessment of the universality of the model.
The structures produced in this fashion are a class of the most challenging configurations for prediction because of their highly disordered nature. 
In fact, structures encountered in practical problems are generally much more stable than disordered structures in terms of energy.

The second type consists of the adsorbed structures. It consists of a crystal surface and small molecules that are sufficiently close to interact.

The third type consists of molecules generated through normal-mode sampling (NMS). Specifically, this refers to the structures of organic molecules containing eight heavy atoms (including C, N, O, P, and S) whose atomic positions fluctuate according to the NMS method.\cite{ani1}

Table \ref{tab:dataset_regression} shows the prediction performances of the energy and force for these components.
For the last two realistic components, we can see that the PFP can predict energy and force with high accuracy. Structures used in this section are not used for the training process.

The detailed scatter plots corresponding to this section is available in Supplementary Data 5.
 
\begin{table}[htbp]
  \centering
  \begin{tabular}{lD{.}{.}{3}D{.}{.}{3}}
Lattice & \multicolumn{1}{c}{Energy MAE} & \multicolumn{1}{c}{Force MAE} \\
 & \multicolumn{1}{c}{[meV/atom]} & \multicolumn{1}{c}{[eV/\angstrom]} \\ \hline
Disordered structure & 13.6 & 0.13 \\
Adsorbed structure & 5.6 & 0.065 \\
Molecule NMS structure (Molecule mode) & 2.6 & 0.034 \\
  \end{tabular}
  \caption{Energy and force regression performance among the datasets.}
  \label{tab:dataset_regression}
\end{table}

\clearpage
\section*{Data 3: Estimated silicon crystal properties using NNPs trained using OC20 dataset} \label{sec:si_silicon_oc20}

Table \ref{tab:oc20_si} shows the estimated relative energies and densities of silicon crystals. ``DimeNet++'' and ``PFP (OC20)'' are trained using the OC20 dataset. For comparison, PFP trained by our dataset is also shown in the ``PFP (ours)'' column. It should be noted that our dataset contains such crystal structures, and therefore the high accuracy of the results of PFP (ours) was expected.

\begin{figure*}[htbp]
\centering
\includegraphics[width=0.80\linewidth]{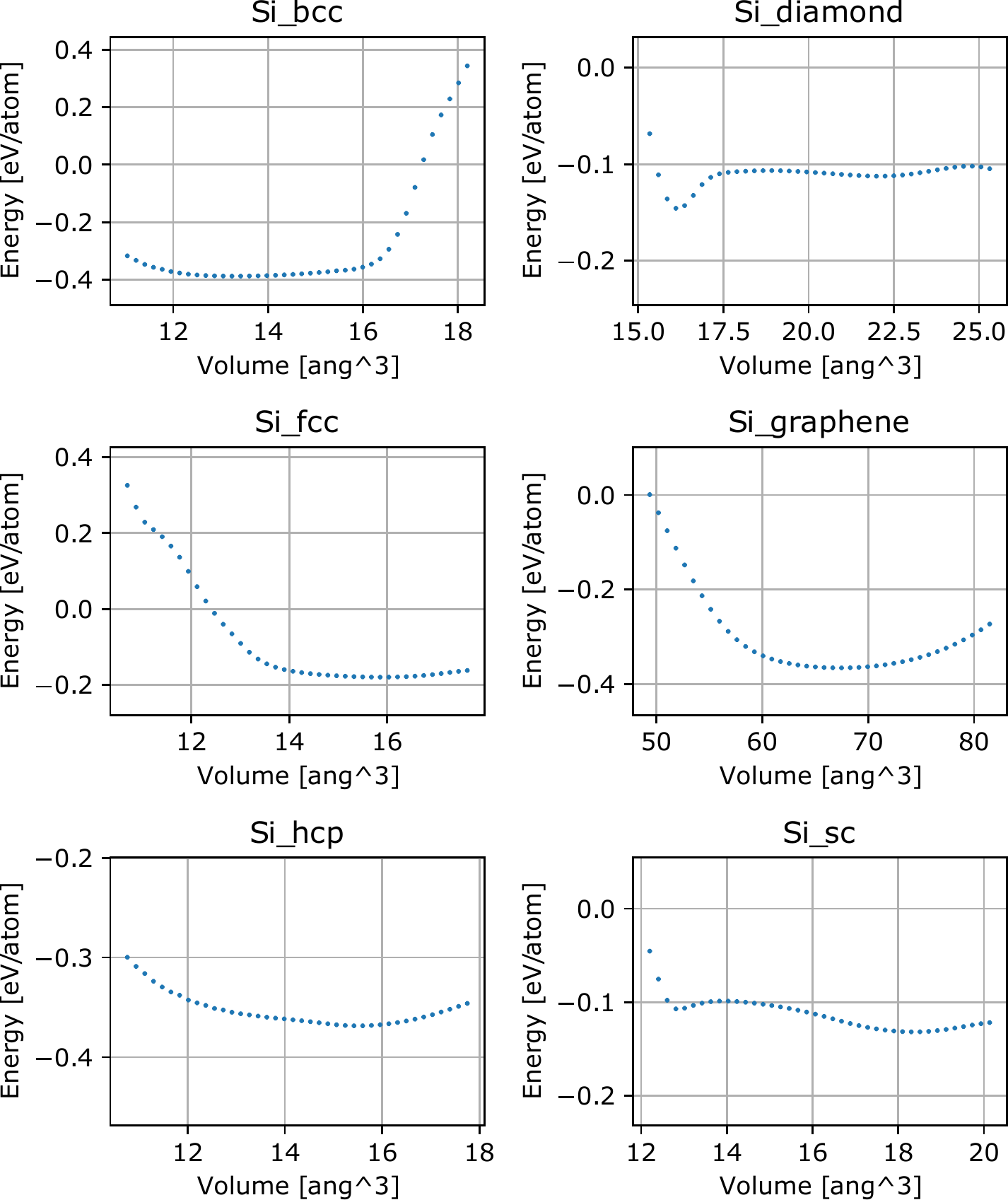}
\caption{Energy curve of DimeNet++. \cite{oc20data}}
\label{fig:si_crystal_oc20_dimenet}
\end{figure*}

\begin{figure*}[htbp]
\centering
\includegraphics[width=0.80\linewidth]{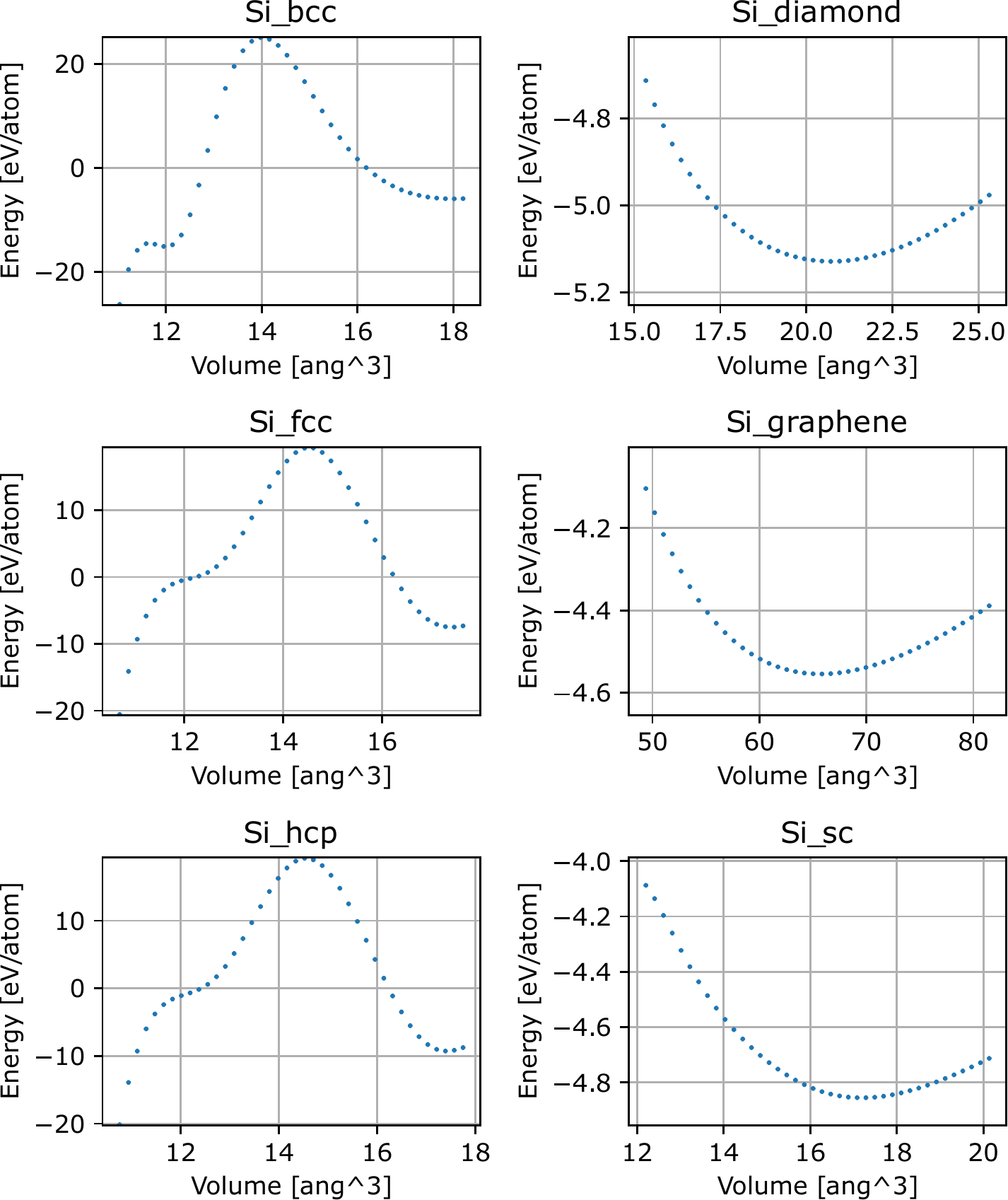}
\caption{Energy curve of PFP architecture trained with OC20 dataset only.}
\label{fig:si_crystal_oc20_pfp_oc20dataset}
\end{figure*}

\begin{figure*}[htbp]
\centering
\includegraphics[width=0.80\linewidth]{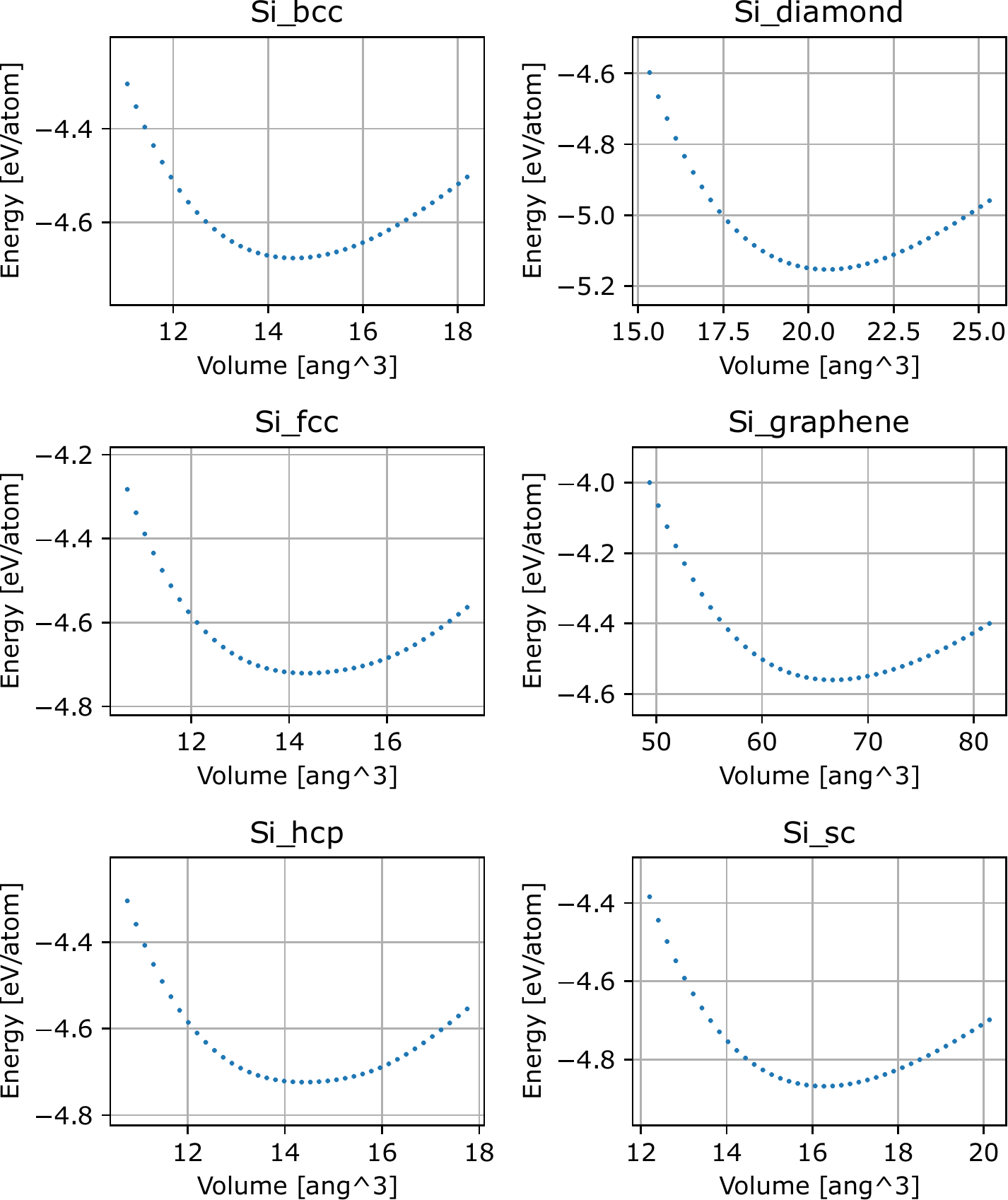}
\caption{Energy curve of ordinary PFP (trained using our dataset).}
\label{fig:si_crystal_oc20_pfp_full}
\end{figure*}

The calculated wide energy surfaces are shown in Figures \ref{fig:si_crystal_oc20_dimenet}, \ref{fig:si_crystal_oc20_pfp_oc20dataset}, and \ref{fig:si_crystal_oc20_pfp_full}.

\begin{table*}[htbp]
  \centering
  \begin{tabular}{llD{.}{.}{4}D{.}{.}{4}D{.}{.}{4}D{.}{.}{4}}
 Property & Lattice & \multicolumn{1}{c}{DFT} & \multicolumn{1}{c}{DimeNet++} & \multicolumn{1}{c}{PFP (OC20)} & \multicolumn{1}{c}{PFP (ours)} \\ \hline
Relative energy    & Diamond & (-4.56) & & & \\
~[eV/atom]          & FCC & 0.54 & -0.07 & -2.36 & 0.43 \\
                   & HCP & 0.49 & -0.26 & -4.09 & 0.43 \\
                   & BCC & 0.55 & -0.28 & -0.85 & 0.48 \\
                   & SC  & 0.32 & -0.02 & 0.26 & 0.29 \\
                   & Graphene & 0.66 & -0.25 & 0.58 & 0.59 \\
Density            & Diamond  & 2.28 & 2.09 & 2.28 & 2.28 \\
~[g/cm${}^3$]       & FCC & 3.27 & 2.92 & 2.68 & 3.23 \\
                   & HCP & 3.25 & 2.99 & 2.66 & 3.25 \\
                   & BCC & 3.17 & 3.56 & 2.60 & 3.21 \\
                   & SC  & 2.87 & 2.58 & 2.71 & 2.87 \\
  \end{tabular}
  \caption{Comparison of estimated relative energies (compared to diamond structure) and densities of silicon crystals.}
  \label{tab:oc20_si}  
\end{table*}

\clearpage
\section*{Data 4: PFP architecture for OC20 task}
For the OC20 task, we essentially used the same architecture as the original PFP. However, the following points have been modified from the original:
The NN parameters derived from the PFP dataset and the corresponding DFT calculations were excluded. This includes a shift in the value of the energy in a vacuum for each element.
Training was conducted using only the OC20 S2EF 2M dataset, and no validation dataset was used during training.
The validation dataset contains one million structures for each task.
During inference, energy was clipped to a maximum of 10.0 eV/atom, and the force was clipped to a maximum of 100.0 eV/\angstrom. There were three, four, three, and seven samples that met the conditions for the ID, OOD ads, OOD cat, and OOD for both tasks, respectively.

\clearpage
\section*{Data 5: Dataset regression performance} \label{sec:si_dataset_regrssion}
\begin{figure*}[htbp]
    \begin{tabular}{cc}
        \centering
        \begin{minipage}[t]{0.40\linewidth}
            \centering
            \includegraphics[width=0.99\linewidth]{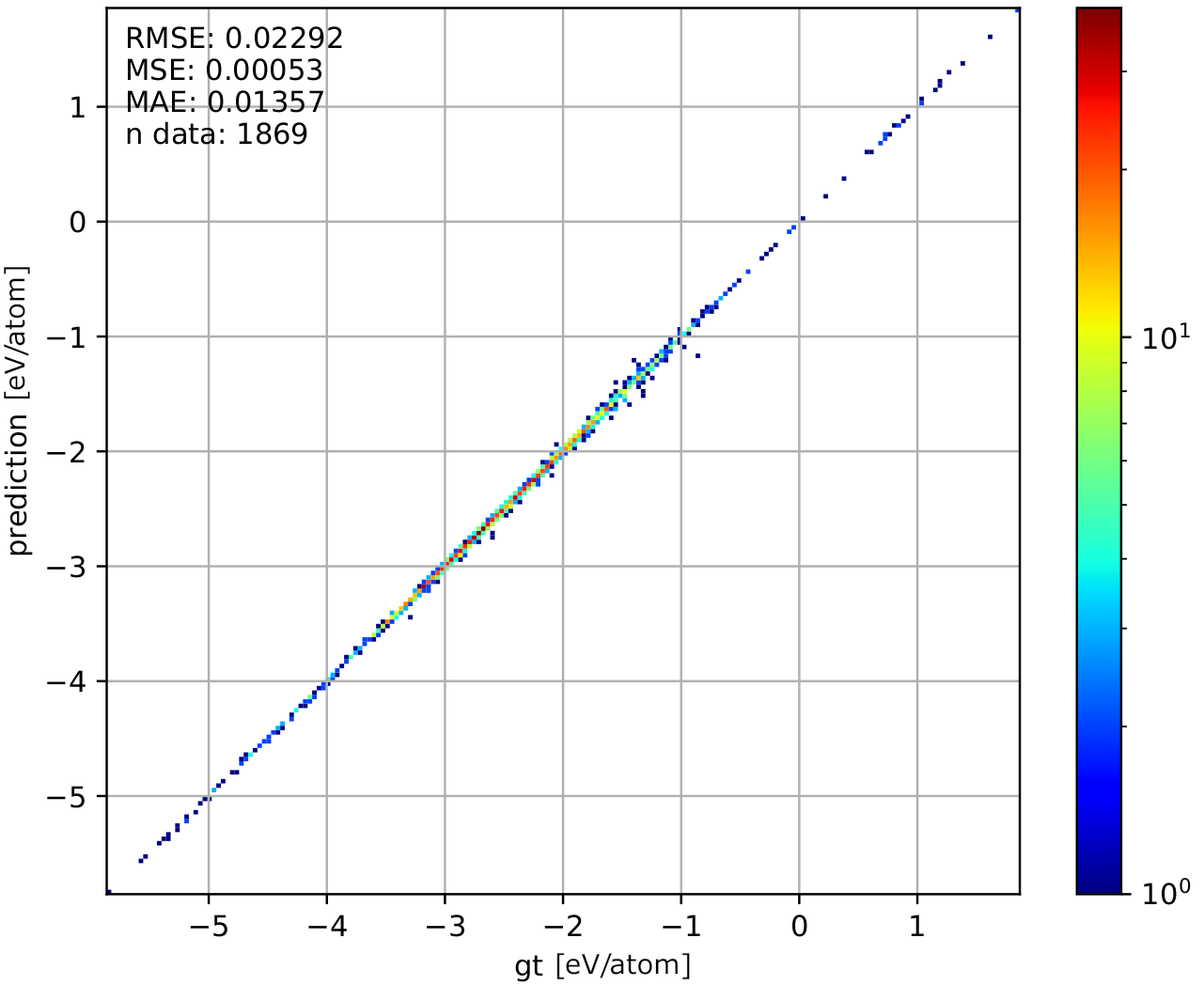}
            (a)
        \end{minipage} &
        \begin{minipage}[t]{0.40\linewidth}
            \centering
            \includegraphics[width=0.99\linewidth]{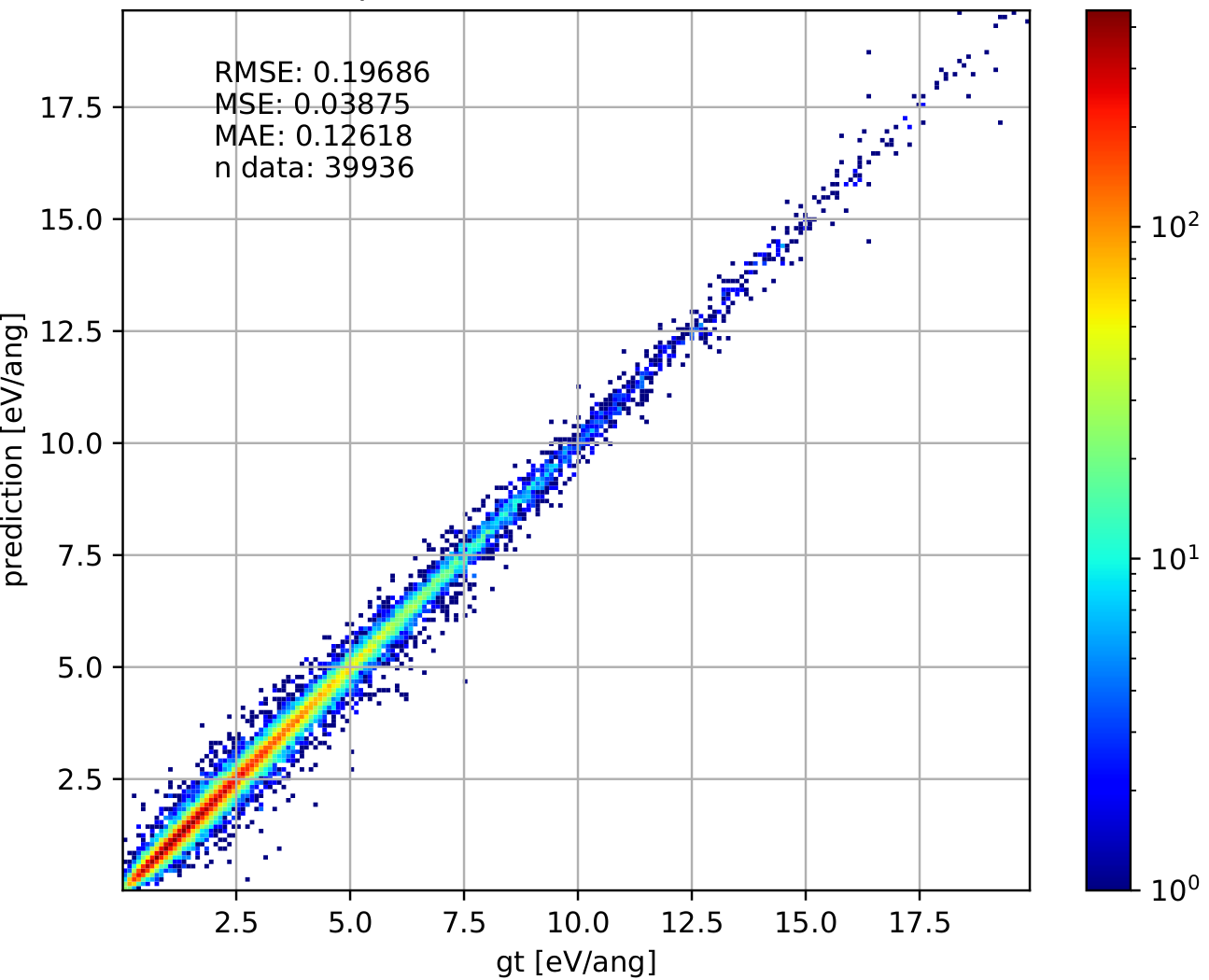}
            (b)
        \end{minipage} \\
        \begin{minipage}[t]{0.40\linewidth}
            \centering
            \includegraphics[width=0.99\linewidth]{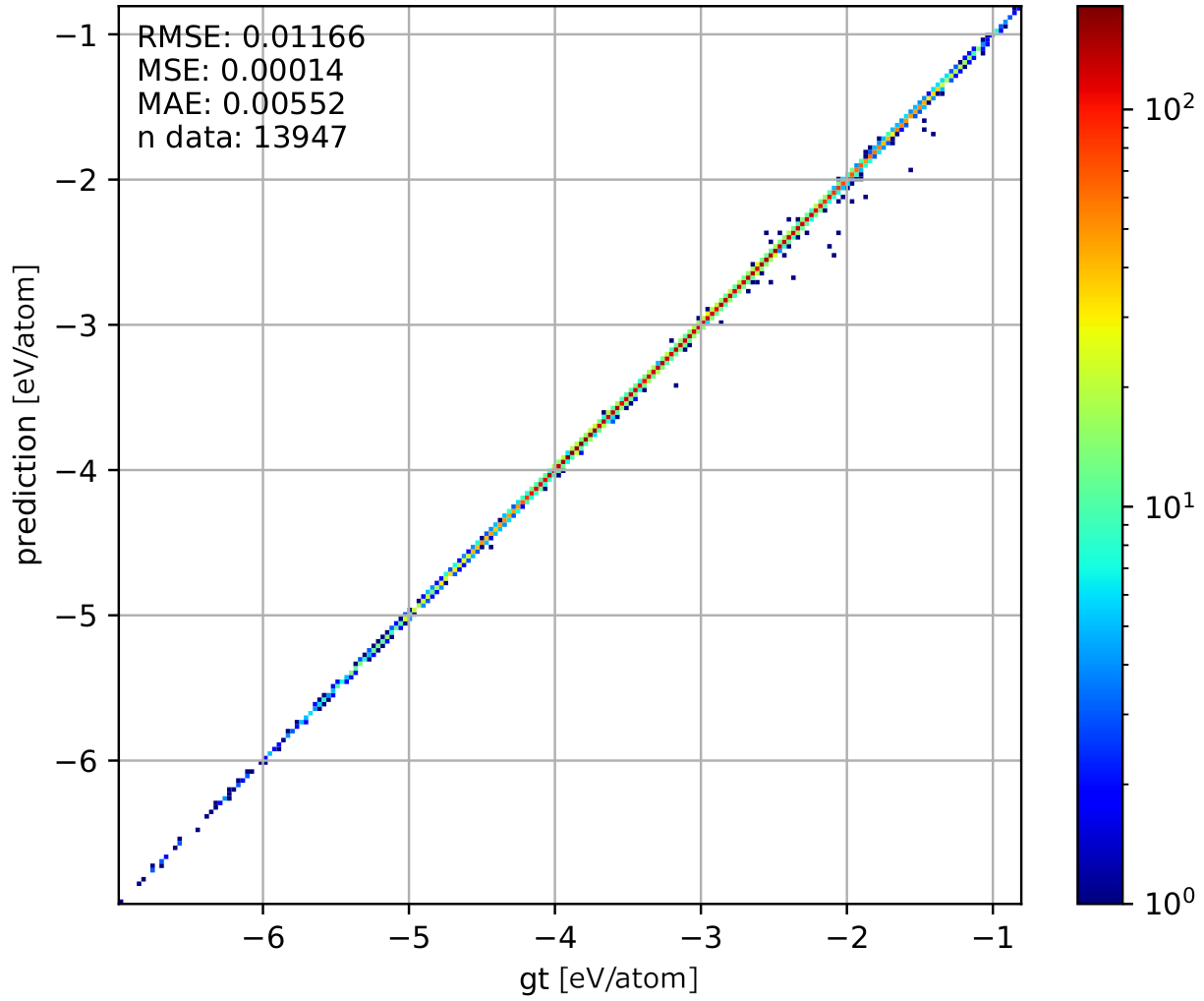}
            (c)
          \end{minipage} &
          \begin{minipage}[t]{0.40\linewidth}
            \centering
            \includegraphics[width=0.99\linewidth]{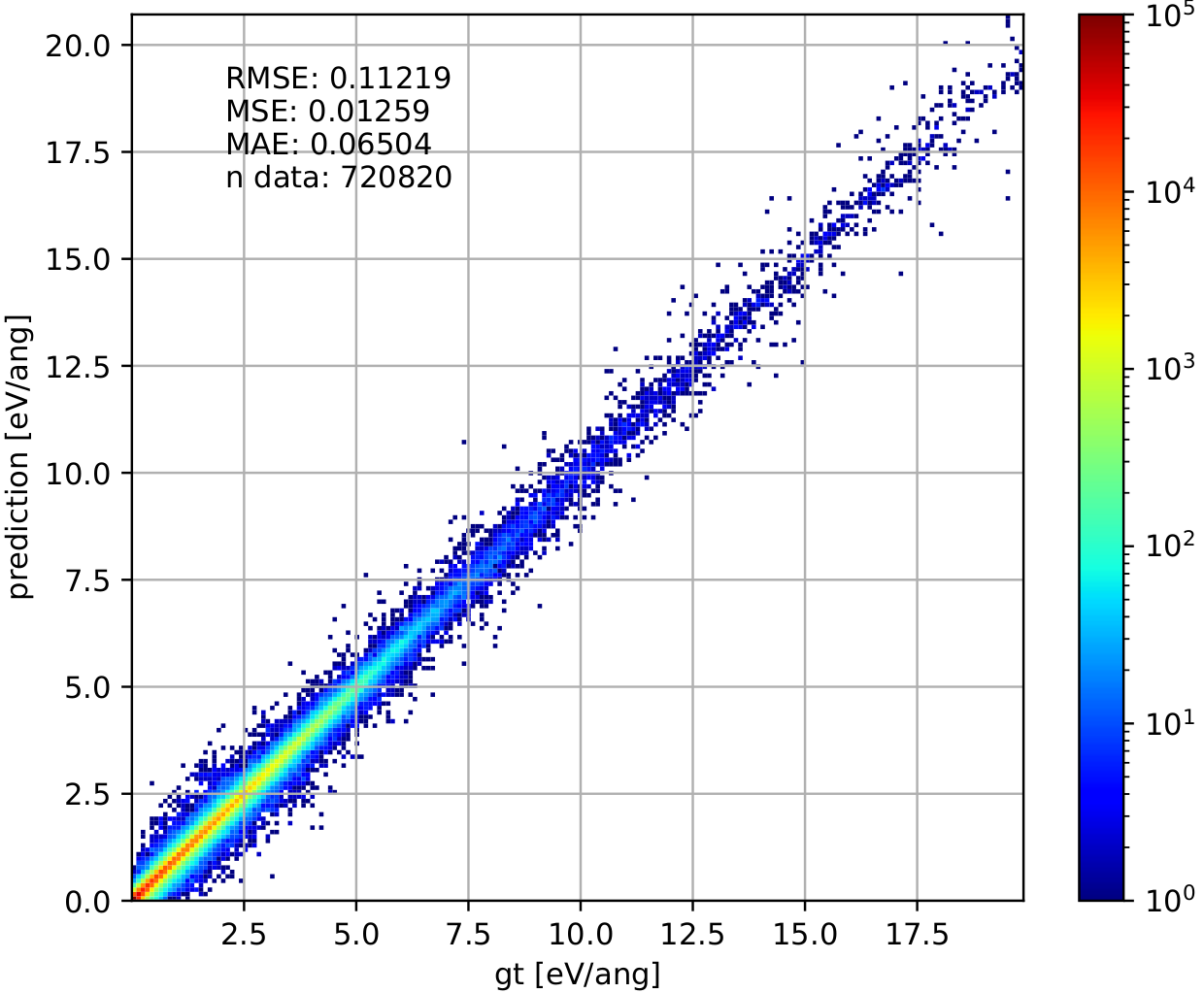}
            (d)
          \end{minipage} \\
        \begin{minipage}[t]{0.40\linewidth}
            \centering
            \includegraphics[width=0.99\linewidth]{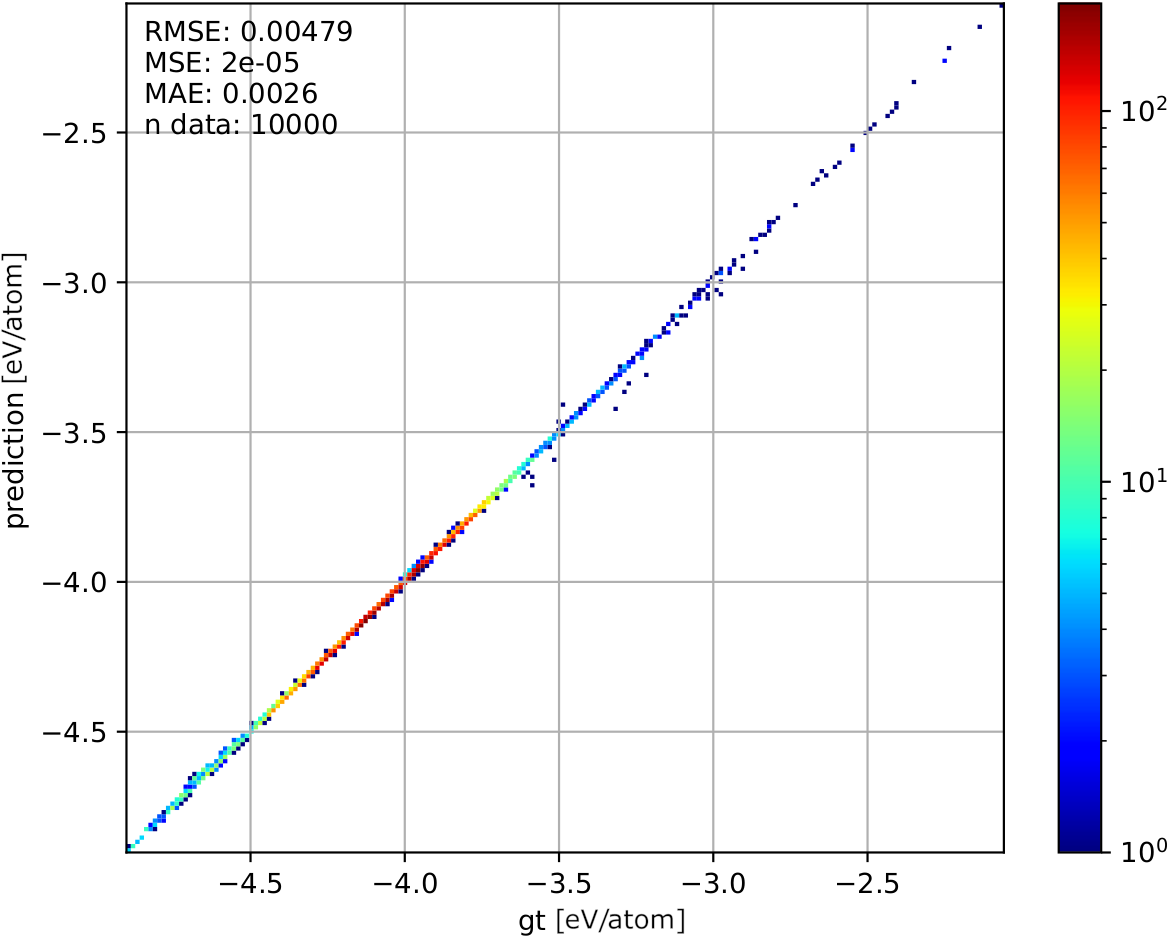}
            (e)
        \end{minipage} &
        \begin{minipage}[t]{0.40\linewidth}
            \centering
            \includegraphics[width=0.99\linewidth]{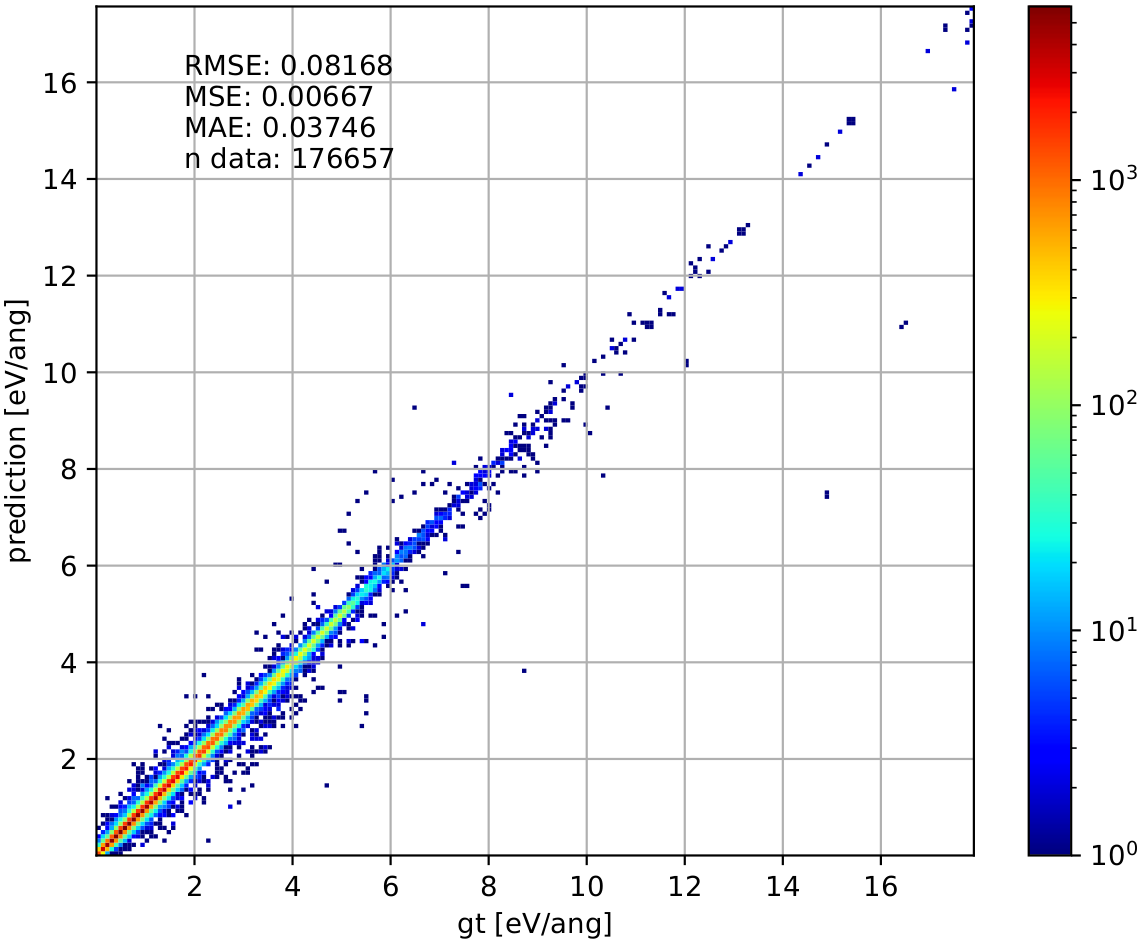}
            (f)
        \end{minipage} \\
    \end{tabular}
\centering
\caption{Energy and force regression performance. The left column corresponds to energy (eV/atom), and the right column corresponds to the force (eV/\angstrom). (a), (b) Disordered structures. (c), (d) Surface adsorbed structures. (e), (f): PubChem molecule normal-mode sampling (NMS) structures.}
\label{fig:dataset_regression}
\end{figure*}

Figure \ref{fig:dataset_regression} shows a scatter plot of the regression results for energy and force in the sub components of the PFP dataset. They correspond to test dataset, which were randomly extracted from the dataset and not used during the training process. See Supplementary Data 10 for the description of each components.

\clearpage
\section*{Data 6: Calculation time benchmark}
The energy and force calculation time for a system of 3000 Pt atoms using PFP was 0.3 s. By contrast, the estimated typical DFT calculation time for the same system is approximately 2 months, which means that PFP is 20-million times faster than DFT.

To estimate the DFT calculation time, we used \textsc{QUANTUM ESPRESSO}\cite{QE-2009,QE-2017} version 6.4.1. The pseudopotential applied for the calculation was \mbox{Pt.pbe-n-kjpaw\_psl.1.0.0.UPF} from \mbox{http://www.quantum-espresso.org}, and the cutoff energy was set to 40 Ry. The calculation time was measured on an Intel Xeon Gold 6254 3.1 GHz$\times$2 (36-core) CPU.
The measured structures were bulk FCC platinum systems with 32, 108, and 256 atoms, and the calculation times were 34, 811, and 8280 s, respectively. We extrapolated the time required for the structure of 3000 atoms by fitting these values to a one-log graph. The fitted line shows that the calculation time is proportional to $O\left(N^{2.64}\right)$, where $N$ is the number of atoms.

The PFP calculation time was measured on an NVIDIA V100 (single GPU).

\clearpage
\section*{Data 7: Structure examples of PFP dataset}
\begin{figure}[htbp]
    \begin{tabular}{ccc}
        \centering
        \begin{minipage}[t]{0.33\linewidth}
            \centering
            \includegraphics[width=0.99\linewidth]{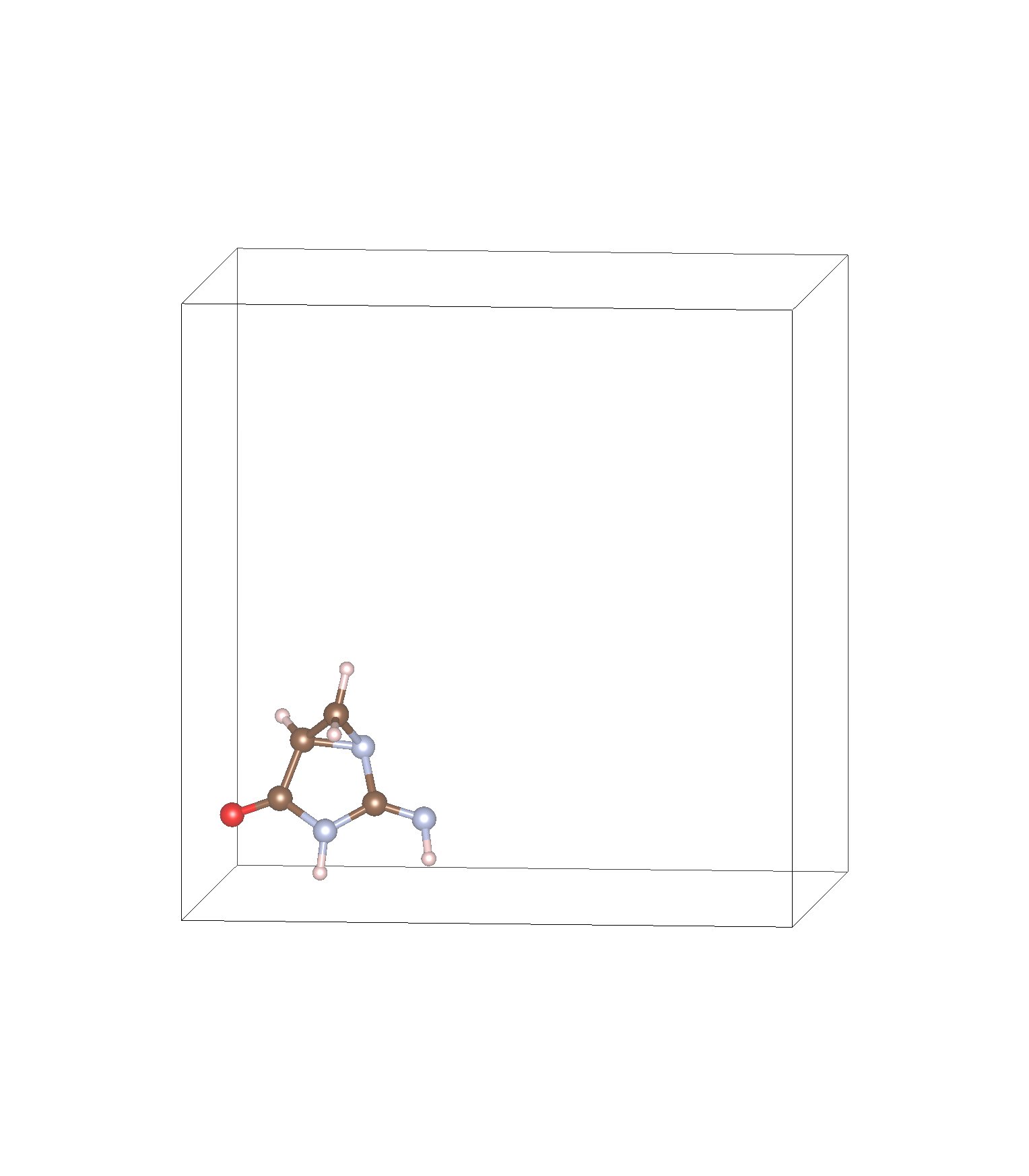}
            (a)
        \end{minipage} &
        \begin{minipage}[t]{0.33\linewidth}
            \centering
            \includegraphics[width=0.99\linewidth]{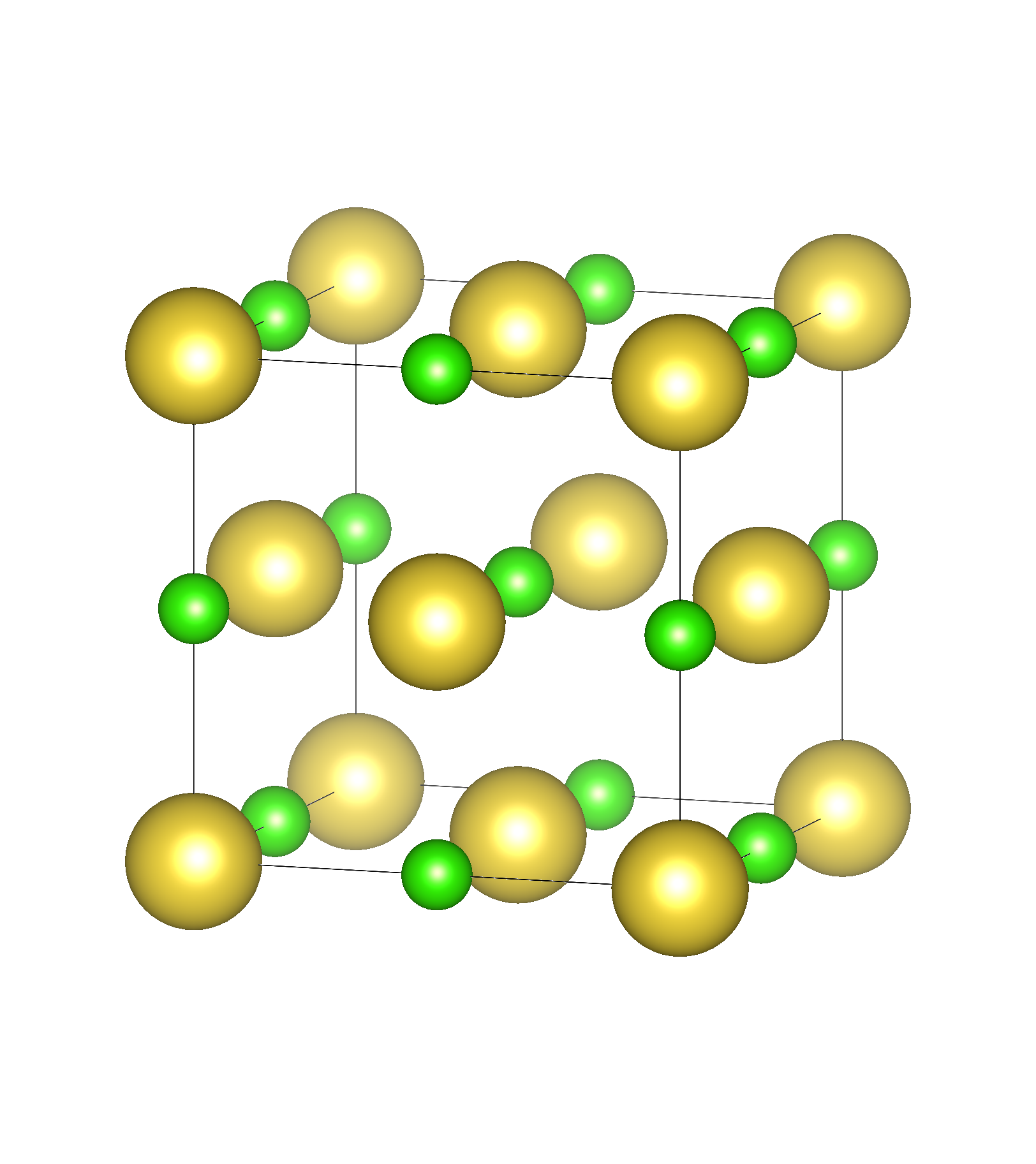}
            (b)
        \end{minipage} &
        \begin{minipage}[t]{0.33\linewidth}
            \centering
            \includegraphics[width=0.99\linewidth]{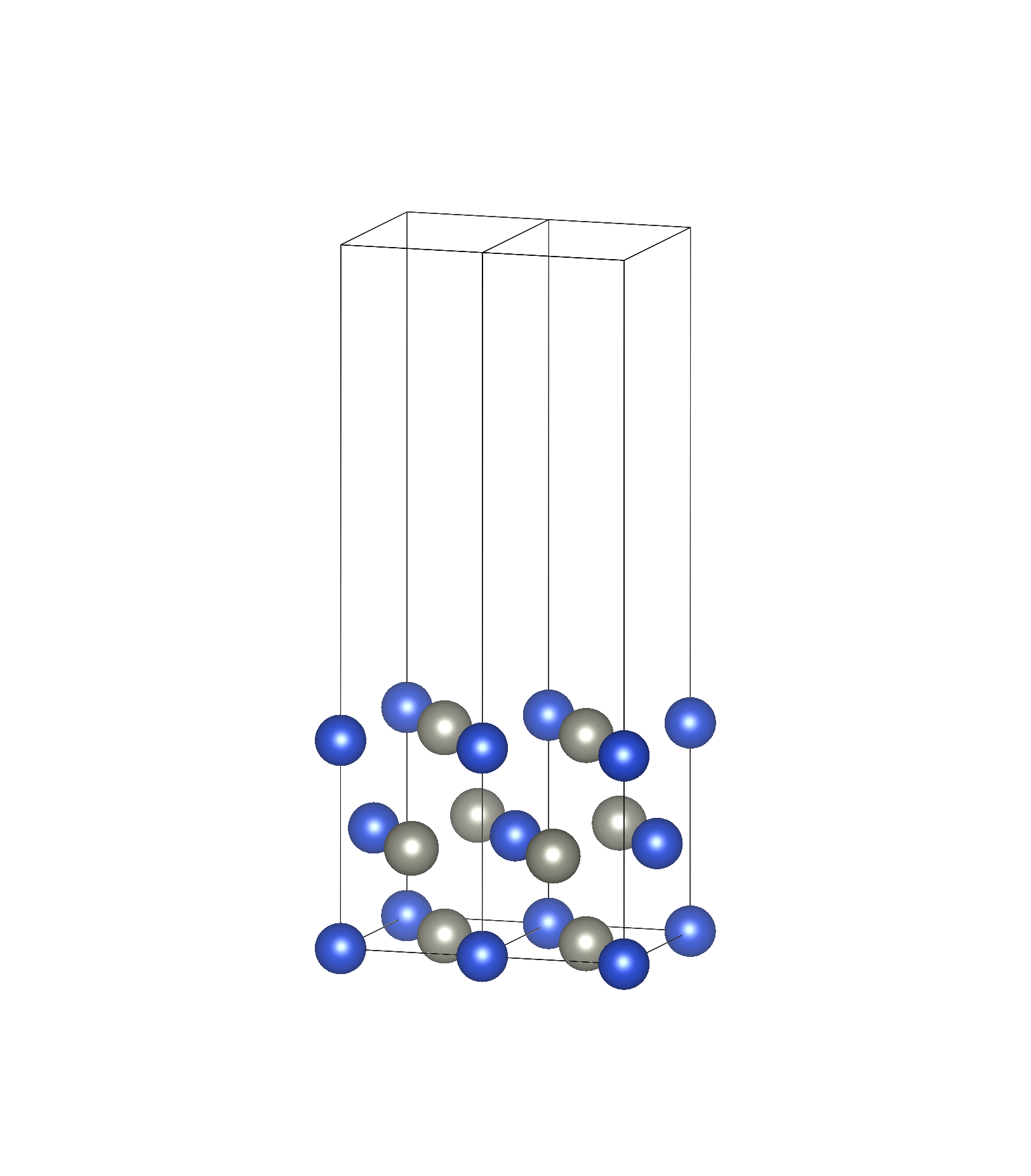}
            (c)
        \end{minipage} \\
        \begin{minipage}[t]{0.33\linewidth}
            \centering
            \includegraphics[width=0.99\linewidth]{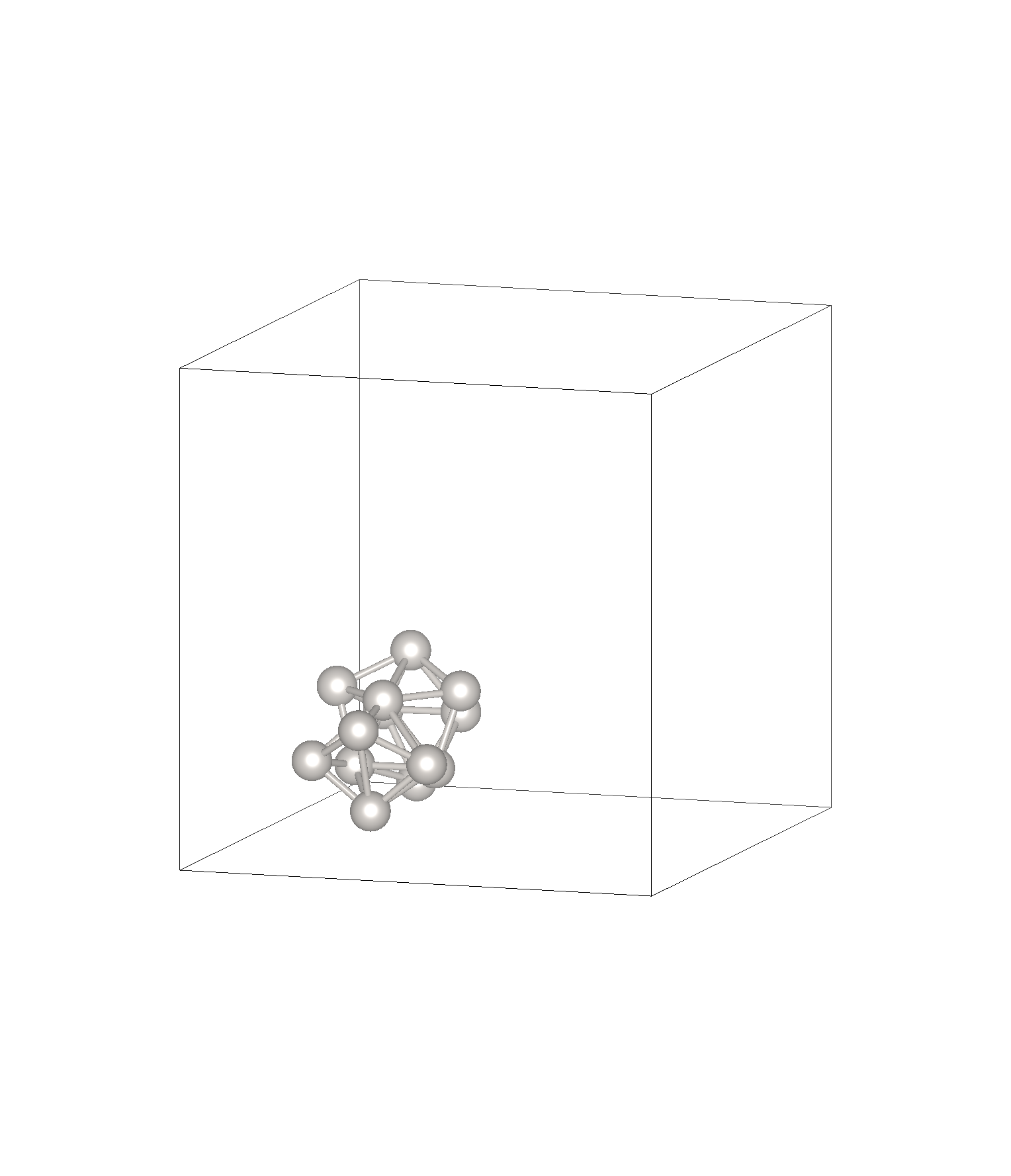}
            (e)
        \end{minipage} &
        \begin{minipage}[t]{0.33\linewidth}
            \centering
            \includegraphics[width=0.99\linewidth]{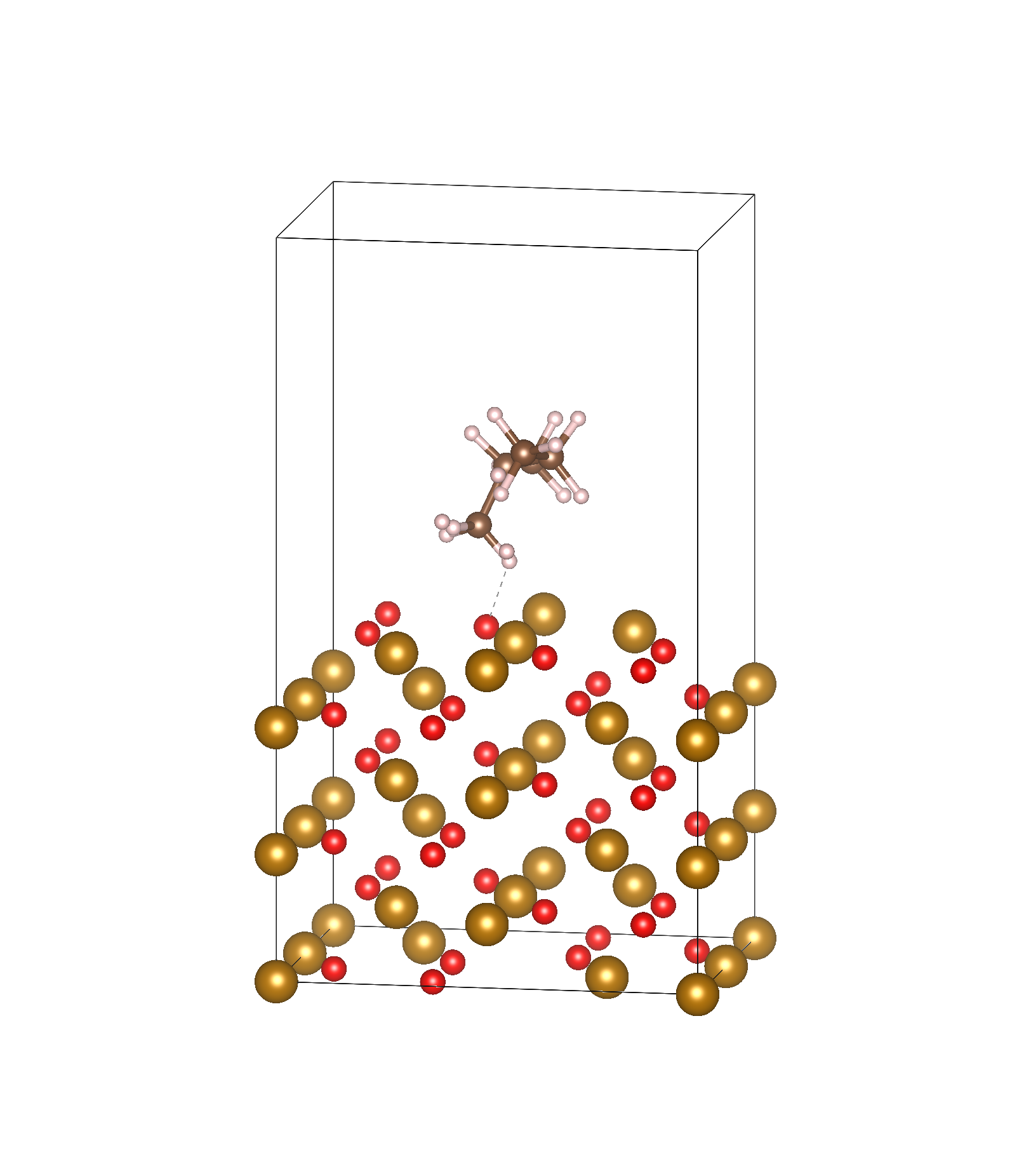}
            (e)
        \end{minipage} &
        \begin{minipage}[t]{0.33\linewidth}
            \centering
            \includegraphics[width=0.99\linewidth]{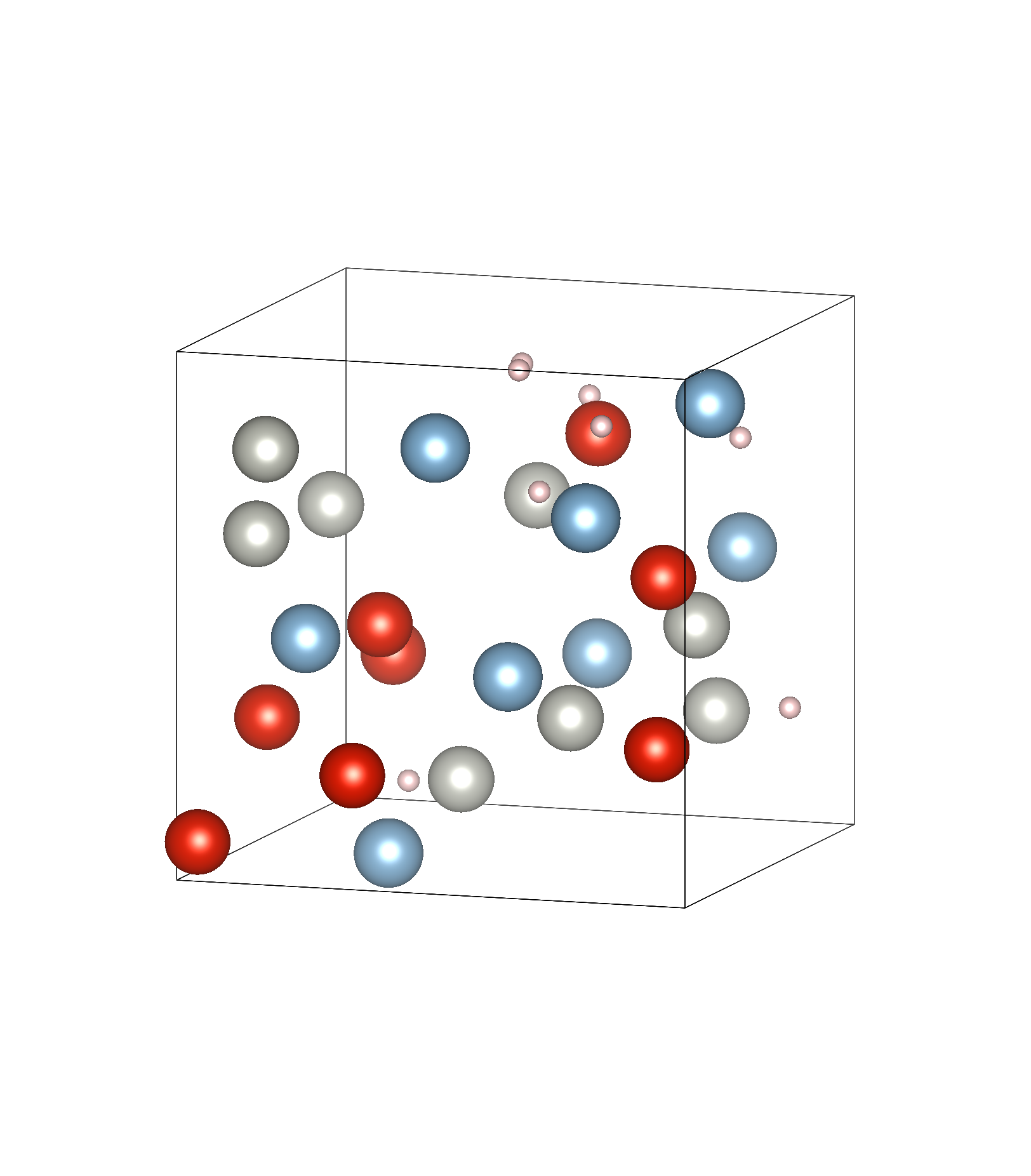}
            (f)
        \end{minipage} \\
    \end{tabular}
    \caption{Structural examples of PFP dataset: (a) molecule, (b) bulk, (c) slab, (d) cluster, (e) adsorption, and (f) disordered are shown.}
    \label{fig:structure_example}
\end{figure}

Figure \ref{fig:structure_example} shows an example of a structure in the PFP dataset.

\clearpage
\section*{Data 8: Elements supported in the OC20 dataset}
\begin{figure}[htbp]
    \centering
    \includegraphics[width=0.99\linewidth]{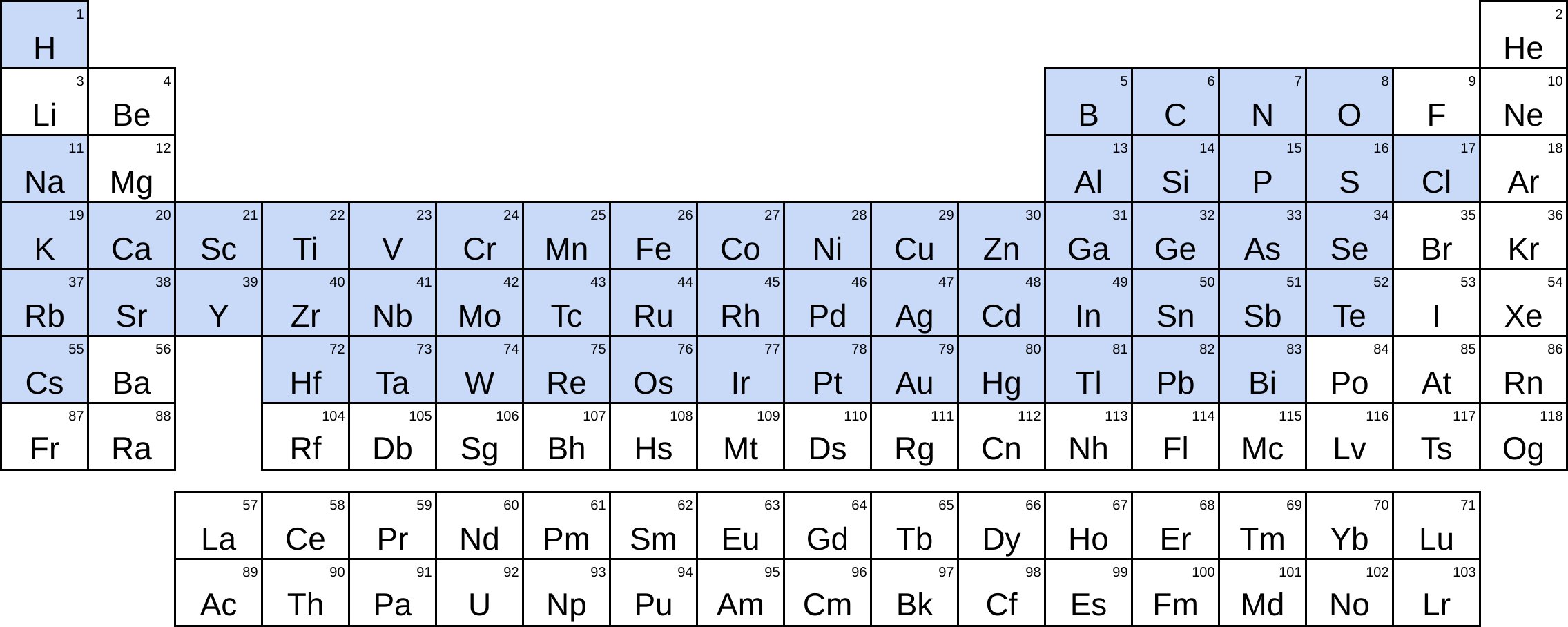}
    \caption{The 56 elements supported by the OC20 dataset are colored in the periodic table.}
    \label{fig:oc20_elements}
\end{figure}

Figure \ref{fig:oc20_elements} shows the 56 elements supported by the OC20 dataset.

\clearpage
\section*{Data 9: NNP architecture and invariances} \label {sec:invariances}
In general, it is essential for machine learning models to incorporate the inductive bias of the target domain to improve the accuracy and generalization. Some of the properties imposed on NNs for atomic structures include rotational invariance, translational invariance, and mirror-image reversal invariance. Among them, the one with rotational invariance is called $\mathrm{SO}(N)$, that with translational invariance in addition to $\mathrm{SO}(N)$ is called $\mathrm{SE}(N)$, and the one with mirror-image inversion invariance in addition to $\mathrm{SE}(N)$ is called $\mathrm{E}(N)$.
When they are not equipped, physically unnatural effects occur, such as unnatural external forces or the inferring of different energies for optical isomers.

However, from the viewpoint of improving the representational performance, there is a demand to design an architecture without losing higher-order features related to positional relationships.
For example, in architectures based on atomic environment vectors (AEV) and node-based machine-learning potentials (MLPs) \cite{bpnn,ani1,ani2x,sch-ani,tensormol}, the positional relationship information is represented by the bond distances and angles.
Rich local positional information is provided while preserving the invariance. However, because the message-passing mechanism of graph neural networks (GNNs) is lacking, positional information farther than the cutoff distance cannot be conveyed.
Another problem is the explosion of a combination of elements. The angle term in the AEV has a number of combinations proportional to the cube of the type of element.

Various methods have been devised to handle higher-order structural information in GNN architectures. In Table \ref{tab:nnp_invariances}, we compare our architecture with previous studies in terms of invariance.
These methods can be broadly classified into two categories: those that use spherical harmonics features and those that use vectors. The former has an invariance of $\mathrm{SE}(3)$, and the latter has an invariance of $\mathrm{E}(3)$. Among these, TeaNet \cite{teanet} has $\mathrm{E}(3)$ invariance and can handle higher-order features such as second-order tensor quantities. Therefore, we adopted the TeaNet-style tensor-based convolution layer for our GNN architecture.

\begin{table*}[hbp]
  \centering
  \begin{tabular}{ccccccc}
Architecture & Invariance & NN type & Scalar & sph & Vector & Tensor \\ \hline
BPNN~\cite{bpnn}               & $E(3)$ Invariant & MLP & \checkmark & & & \\
ANI-1~\cite{ani1}              & $E(3)$ Invariant & MLP & \checkmark & & & \\
ANI-2x~\cite{ani2x}              & $E(3)$ Invariant & MLP & \checkmark & & & \\
Schrodinger-ANI~\cite{sch-ani} & $E(3)$ Invariant & MLP & \checkmark & & & \\
TensorMol-0.1~\cite{tensormol}      & $E(3)$ Invariant & MLP & \checkmark & & & \\
SchNet~\cite{schnet2017} & $E(3)$ Invariant & GNN & \checkmark & & & \\
DimeNet++~\cite{dimenet,dimenetpp} & $E(3)$ Invariant & GNN & \checkmark & & & \\
PhysNet~\cite{physnet} & $E(3)$ Invariant & GNN & \checkmark & & & \\
Cormorant~\cite{cormorant} & $SE(3)$ Equivariant & GNN & \checkmark & \checkmark & & \\
SE(3)-Transformer~\cite{se3tr} & $SE(3)$ Equivariant & GNN & \checkmark & \checkmark & & \\
NequIP~\cite{nequip} & $SE(3)$ Equivariant & GNN & \checkmark & \checkmark & & \\
SpookyNet~\cite{spookynet} & $SE(3)$ Equivariant & GNN & \checkmark & \checkmark & & \\
EGNN~\cite{egnn} & $E(3)$ Equivariant & GNN & \checkmark & & \checkmark & \\
TeaNet~\cite{teanet} & $E(3)$ Equivariant & GNN & \checkmark & & \checkmark & \checkmark \\
ours & $E(3)$ Equivariant & GNN & \checkmark & & \checkmark & \checkmark \\
  \end{tabular}
  \caption{Categorization of recent NNP architectures. Here, ``scalar'' denotes a rotation invariant feature, also often called an atomic environment vector (AEV) when used with MLPs. In addition, ``sph'' denotes a higher-order spherical harmonics feature, and ``vector'' and ``tensor'' represent first- and second-order rotational equivariant features, respectively. $\mathrm{SE}(3)$ models can also be turned into an $\mathrm{E}(3)$ model with additional constraints.}
  \label{tab:nnp_invariances}
\end{table*}

Note that when inputting a graph structure into an NN, nodes are transformed once into an ordered list, and NNPs generally need to also satisfy permutation invariance for the order of the nodes. All of the above models, including ours, satisfy this permutation invariance.

\clearpage
\section*{Data 10: Structure generation details of PFP dataset}
\subsection*{molecule}
Base molecules were obtained from GDB-11 datasets \cite{fink2007virtual,fink2005virtual} or PubChem databases.\cite{10.1093/nar/gkaa892} Additional base molecules were generated by modifying the elements of the base molecules. For example, diatomic molecules for all pairs of 45 elements were generated to obtain diatomic potentials.
The structures of the molecules are generated by geometrical optimization, normal mode sampling (NMS) \cite{ani1}, or molecular dynamics. Structures with two molecules are also generated with molecular dynamics. The settings of NMS are approximately the same as those of the original method, but the number of structures to be sampled is changed to reduce the computational cost.

\subsection*{bulk}
For the base structures of the bulk material, we gathered various single-element and binary-element crystal base structures with one or two elements for all 45 elements or pairs of 45 elements.
For single-element crystal base structures, simple cubic, face-centered cubic, body-centered cubic, hexagonal close-packed, and diamond structures were prepared.
One-dimensional chain structures and two-dimensional graphene structures were also prepared as single-element crystal base structures.
For binary-element crystal base structures, zincblende, rocksalt, cesium chloride, and wurtzite structures were prepared.
Furthermore, additional base structures were obtained from the materials project database \cite{MaterialsProject} under the condition that the number of elements is two, the number of sites is less than or equal to 16, and at least one transition metal element (Ti – Zn, Mo, Ru – Ag) is included.
The above base structures were volume-relaxed with DFT while maintaining the cell shapes and fractional coordinates of each site.
Subsequently, additional deformed structures were generated for each base structure based on the volume-relaxed structure in four ways.

\subsubsection*{cell compression / expansion}
We sampled 20 deformed structures for each base structure with this method.
The cell volume was compressed or expanded while maintaining the cell shape and fractional coordinates of each site.
In addition, 10 expanded structures were sampled for each base structure, where the relative cell length varied between 1.0 and 1.1.
Another 10 structures were also sampled with lengths between 1.0 and $1.0/1.1$.

\subsubsection*{site position displacement}
For each base structure, 10 deformed structures were sampled with this method.
The Cartesian coordinates of each site were randomly displaced while maintaining the cell shape and volume. 
The displacement distances of each site were obtained from the uniform distribution of 0 to 0.1 × (average distance between sites), and the displacement directions of each site were uniformly sampled.

\subsubsection*{cell shear deformation}
Up to 60 deformed structures were sampled for each base structure with this method.
Shear deformation was applied while maintaining the cell volume and the fractional coordinates of each site.
A single parameter was selected from the lattice angle parameters $\alpha$, $\beta$, and $\gamma$, while the other two were fixed. 
For each base structure, 21 structures were sampled by changing the selected angle parameter from $-5 \degree$ to $+5 \degree$ at even intervals. 
By removing the original structure, 20 structures were obtained.
The corresponding shear plane was scaled to maintain the area.
For example, in the case where the $\alpha$-angle is selected, the b- and c-axis lengths are changed to maintain the area.
Since there are three lattice angle parameters, $20 \times 3 = 60$ structures were generated from a one-volume relaxed structure.
To reduce the calculation cost, symmetrically unique structures were selected from the 60 structures for the calculations using DFT.

\subsubsection*{cell tensile deformation}
Up to 60 deformed structures were sampled for each base structure with this method.
Tensile deformation was applied while maintaining the cell volume, cell angles, and fractional coordinates of each site.
One of the a-, b-, or c-axis lengths was selected and scaled by arithmetic progression from 0.9 to 1.1, and 21 structures were obtained for each selected axis.
By removing the original structure, 20 structures were obtained.
The other unselected two axis lengths are also scaled to keep the cell volume.
Symmetrically unique structures selected from 60 structures were calculated with DFT.

\subsection*{cluster}
Small atomic clusters with one element were created based on 13-atom metal clusters.\cite{PhysRevA.79.043202,PhysRevA.79.043203} Covalent radii were used for the interatomic distance. The length of the vacuum region was 10 \angstrom\ along each axis.
To generate deformed structures, ``cell compression / expression'' method was applied to the cluster structures. The length of the vacuum region was kept constant during the deformation.

\subsection*{disordered}
Structures that are far from stable configurations were sampled using molecular dynamics (MD) simulations at high temperatures. These structures are referred to as disordered structures.
An interatomic potential is required to run MD simulations. To overcome this issue, we first created an early stage of PFP without the MD-origin structures and then used it to sample disordered structures. The calculation flow is as follows:
First, the initial structures are created. A cubic cell is used for the initial structure. The typical cell length is 8.8 \angstrom, but smaller cells with 8.4 \angstrom\ and 8.0 \angstrom\ are also prepared. The atoms are added in the form of a $2\times 2\times 2$ face-centered cubic structure (32 atoms) but with different element types. Multiple element selection strategies have been proposed. One method determines each atom type randomly. The maximum element type limit was set to 20. Another method selects 2, 4, or 8 element types repeatedly and feeds them into the cell. In this case, the number of atoms corresponding to each element was changed in units of 4 atoms. For example, the most unbalanced structure is 28 atoms in a certain element and 4 atoms in another element. The last method is similar to the previous one, but up to 24 atoms are eliminated from the structure. This means that the density of the structure is lower.
Then, high-temperature molecular dynamics simulations were carried out. To unify the timescale of the atoms, the masses of all atoms were set to 5 (atomic mass unit). For the first step, the initial temperature was set to 6000 K, and then 1 ps of NPT ensemble at 10000 K was conducted. The second and third steps were 1 ps of the NPT ensemble at 2000 K and 500 K, respectively. Finally, to obtain the localized structure, the vacuum region was created by expanding one axis of the cell by a factor of two without modifying the atom positions, and 1 ps of NPT ensemble at 300 K was carried out. Four snapshots corresponding to the final state of the four processes were extracted and used for the disordered dataset structure.
After DFT calculation, if the maximum atomic force is larger than 20 eV/\angstrom, the structure is discarded.

\subsection*{slab}
Slab structures with (111), (110), (101), (011), (100), (010), and (001) surfaces were generated from volume-relaxed bulk structures. The length of the vacuum region between the surfaces was 10 \angstrom. The number of equivalent layers of the slab is typically 3. If the thickness of the 3-layer slab is too small or large, a larger or smaller number of layers was used, respectively. The site positions and cell parameters were not relaxed.
Deformed structures were also generated from these slab structures. Four deformation methods introduced for bulk systems were applied to the slab structures. The number of generated deformed structures is halved when the computation cost is high. The length of the vacuum region was kept constant for the deformed structures.

\subsection*{adsorption}
Adsorbed structures were generated for pairs of randomly selected slab structures and randomly selected molecules using the following two methods.

\subsubsection*{geometrical optimization}
First, a randomly rotated molecule was placed at random positions in the vacuum region of the slab structure. Adoption sites, such as on-top, bridge, or hollow sites, were not considered when placing the molecule. Then, the adsorbed structure was geometrically optimized in the early stage of PFP while maintaining the cell parameters and site positions of all slab sites, including the sites on the surface. The definition of the early stage of PFP is described in the disordered section. Slab structures without deformation were selected for this method.

\subsubsection*{random placement}
Initially, a randomly rotated molecule was placed at random positions in the vacuum region of the slab structure. Adoption sites, such as on-top, bridge, or hollow sites, were not considered when placing the molecule.
Here, the closest atom is defined as the atom with the smallest distance normalized by its atomic radius. The normalized smallest distance $d_{min}$ was calculated as follows:
$ d_{min} = \min_{i,j}( D_{ij} / (r_i + r_j) ) $, 
where $i$ is the index of the slab atom; $j$ is the index of the molecule atom; $D_{ij}$ is the Cartesian distance between slab atom $i$ and molecule atom $j$; $r_i$ is the atomic radius of atom $i$; and, $r_j$ is the atomic radius of atom $j$. The atomic radii used were determined by Slater\cite{slater1964atomic} and are available on the Mendeleev package.\cite{mendeleev2014} 
Then, the molecule is moved perpendicular to the slab surface so that the smallest normalized distance is the random value obtained from a uniform distribution in the range of 0.8 to 1.3.
Slab structures without deformations or deformed slab structures with cell expansion/compression methods were selected for this method.

\clearpage
\section*{Data 11: Relationship between PFP applications and dataset}
In this section, the correspondence between PFP applications shown in the results section and the dataset is demonstrated.

\subsection{Lithium diffusion}
The crystal structures of LiFeSO4F and FeSO4F are not explicitly included in the dataset. The crystal structures collected in the dataset (``bulk'' section in Supplementary Data 10) are single-element and binary-element systems. This means that there are no three-body interactions of the three different elements in the crystal structure dataset. The nearest structure is likely included in the disordered dataset. It contains multiple types of elements, and there is a chance of having similar local configurations of FeSO4F or LiFeSO4F. The prerequisite knowledge of LiFeSO4F and FeSO4F was not used in the dataset collection phase.

\subsection{Molecular adsorption in metal-organic framework}
Although MOF structures shown in the results section are not explicitly included in the dataset, artificial molecule structures that contain non-organic elements are included in the dataset (``molecule'' section in Supplementary Data 10). We believe that this is one of the structures closest to the dataset. Conversely, it only contains intramolecular interactions and does not have inter-molecular interactions, which have a major impact on the structure parameters of MOF. In addition, water molecule interactions were not explicitly included in the dataset. These intermolecular interactions may be observed in the adsorption dataset described in Supplementary Data 10.

\subsection{Cu-Au alloy order-disorder transition}
The ordered crystal lattices of the Cu-Au alloy are included in the dataset. Deformed and displaced structures are also included. However, the disordered crystal lattice observed in the simulation was not explicitly included in the dataset.

\subsection{Material discovery for a Fischer--Tropsch catalyst}
The potential curve of the CO molecule is also included in the dataset. The adsorption dataset and external OC20 dataset contain adsorbed structures. However, they correspond to the adsorption energy, instead of the bond-breaking process of the molecules. The catalyst effect of the surfaces shown in the results section was not explicitly included in the dataset.

\clearpage
\section*{Data 12: PFP appications comparision with OC20 DimeNet++ model}
To check the performance of the PFP shown in the results section, the results are compared with those of the existing NNP. The Open Catalyst Project baseline model is used for this task because it was trained on a recently proposed large-scale dataset and applies to multi-element systems.

The publicly available trained model ``dimenetpp\_all’’ is available from the Open Catalyst Project implementation (https://github.com/Open-Catalyst-Project/ocp/tree/v0.0.3). This model was trained with the DimeNet++\cite{dimenetpp} architecture on all data from the S2EF task\cite{oc20data} included in OC20.

In this study, two modifications were made to the dimenetpp\_all inference implementation. First, the cell input was modified to ensure differentiability to compute the stress tensor with automatic differentiation to subsequently use it for cell structure optimization. Second, the upper limit of the number of neighbors considered in the GNN was increased from 50 to 150 to avoid losing the continuity of the energy landscape. We noticed that the estimated energy fluctuated during the structural optimization process when the maximum number of neighbors was 50, which is the default setting used for the training process. It was confirmed that increasing the maximum number of neighbors up to 150 suppresses this energy fluctuation, and the structural optimization converges successfully.

The functional use for the DFT calculations is different. For PFP, the dataset is collected based on the PBE functional and compared with previous studies in the main text. Conversely, the OC20 dataset is calculated using the RPBE functional.\cite{rpbe} A direct comparison of the results requires careful consideration. 

\subsection*{Lithium diffusion}
\begin{figure*}[htbp]
\centering
\includegraphics[width=0.90\linewidth]{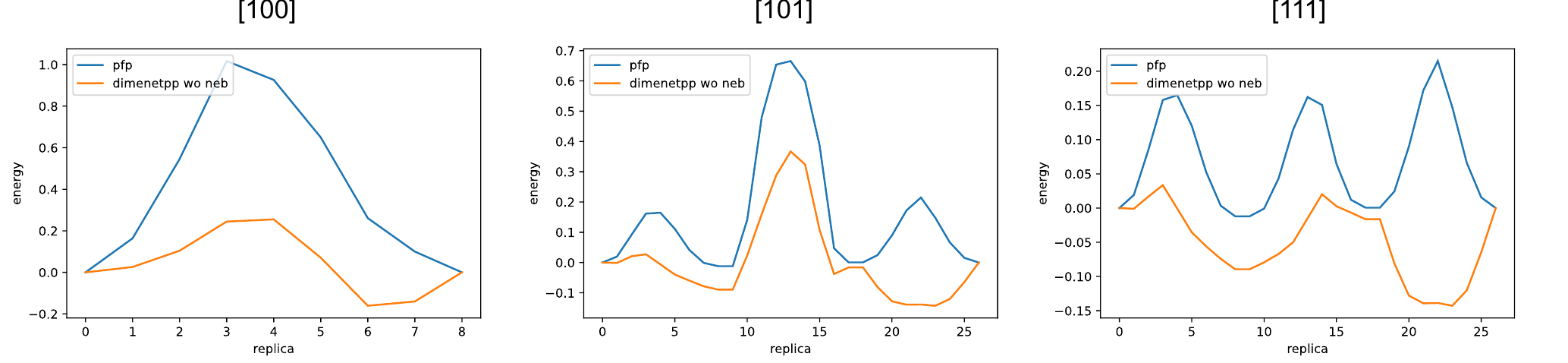}
\caption{NEB calculation images v.s. energies by PFP (blue line). All images were evaluated using dimenetpp\_all (orange line).}
\label{fig:Li_compare}
\end{figure*}

It is not possible to optimize the structure of FeSO${}_4$F using dimenetpp\_all because of energy fluctuations. Therefore, we compared the energy for the same NEB images obtained using PFP. The energy in the initial structure was set to zero in both models.
The results are presented in Fig. \ref{fig:Li_compare}; dimenetpp\_all sometimes fails to reproduce the existence of an energy barrier. Notably, the lithium diffusion phenomenon in the bulk structure is outside the scope of OC20 tasks.

\subsection*{metal-organic framework}
Since the structural optimization does not converge, neither the volume change rate nor the adsorption energy could be evaluated by dimenetpp\_all.

\subsection*{Cu-Au alloy}

It was determined that the bulk energy evaluated by dimenetpp\_all is not appropriate for reproducing the transition phenomena. For example, the formation energy of the CuAu rocksalt structure was quite small ($10^{-7}$ eV/atom order), which is not consistent with the DFT calculation (0.1 eV/atom order, using the PBE functional).

\subsection*{Fischer--Tropsch catalyst}
 NEB calculations were performed using dimenetpp\_all. The same optimization method and NEB conditions were applied. Although the energy values of PFP and dimenetpp\_all differ, the diagrams are relatively reasonable and can be considered an interpolation region of the OC20 dataset.

\begin{figure*}[htbp]
\centering
\includegraphics[width=0.90\linewidth]{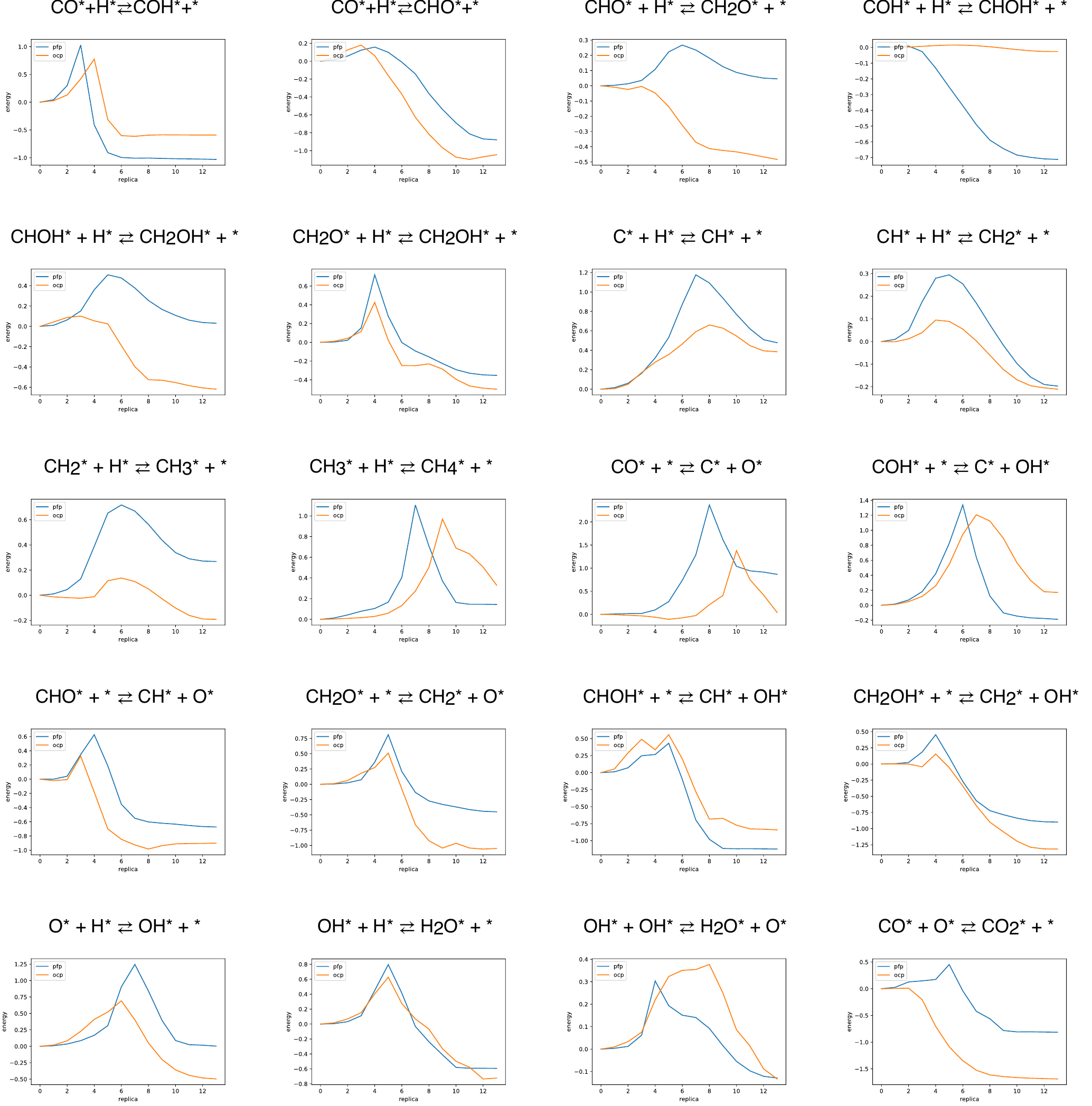}
\caption{Comparison of the NEB calculation result. The blue and orange lines correspond to PFP and dimenetpp\_all, respectively. The energy in the initial state was set to zero.}
\label{fig:FT_compare}
\end{figure*}

\begin{figure*}[htbp]
\centering
\includegraphics[width=0.90\linewidth]{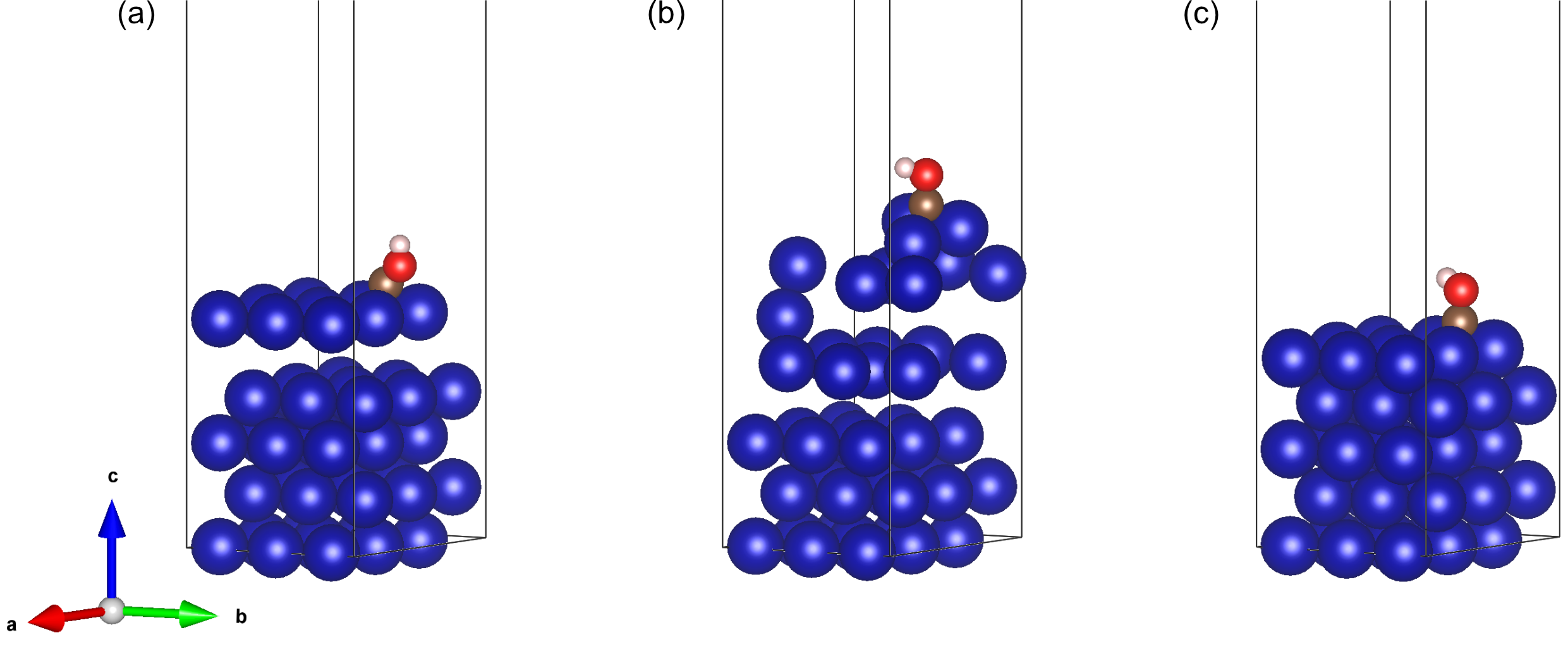}
\caption{Adsorbed structures created by dimenetpp\_all. (a) COH molecule attached to the optimized Co surface (b) Optimized structure of (a). (c) Structure optimized from the COH molecule placed on the Co surface. The figures were drawn using the VESTA visualization package.\cite{momma2011vesta}}
\label{fig:FT_structure}
\end{figure*}

It was difficult to reproduce the adsorption process using dimenetpp\_all. To create an adsorbed structure, the molecule was attached to the optimized surface (Fig. \ref{fig:FT_structure} (a)). However, the structure obtained is not reasonable (Fig. \ref{fig:FT_structure} (b)). Although the adsorbed structure can be obtained by attaching the molecule to the bare surface, where both of them can be optimized simultaneously (Fig. \ref{fig:FT_structure} (c)), the evaluated energy indicates that structure b (-6.15 eV) is more stable than structure c (-1.56 eV).

\clearpage
\section*{Data 13: Molecular dynamics simulation of lithium diffusion}
The detail of the calculation method for the activation energy of lithium diffusion using molecular dynamics simulation is shown below.

First, the same structure as the initial state of the NEB calculation was prepared. Second, the initial momenta were applied based on the Maxwell-Boltzmann distribution. Then, 100 ps of the NVT ensemble was applied at a constant temperature. The temperature was set at 300 K, 325 K, 350 K, 375 K, and 400 K. Eight independent trajectories were sampled for each temperature by changing the random seed of the initial momenta. It was verified that the lithium atom only travels along the $[111]$ direction, which is considered the lowest activation energy path for the CI-NEB calculations.

\begin{figure*}[htbp]
\centering
\includegraphics[width=0.90\linewidth]{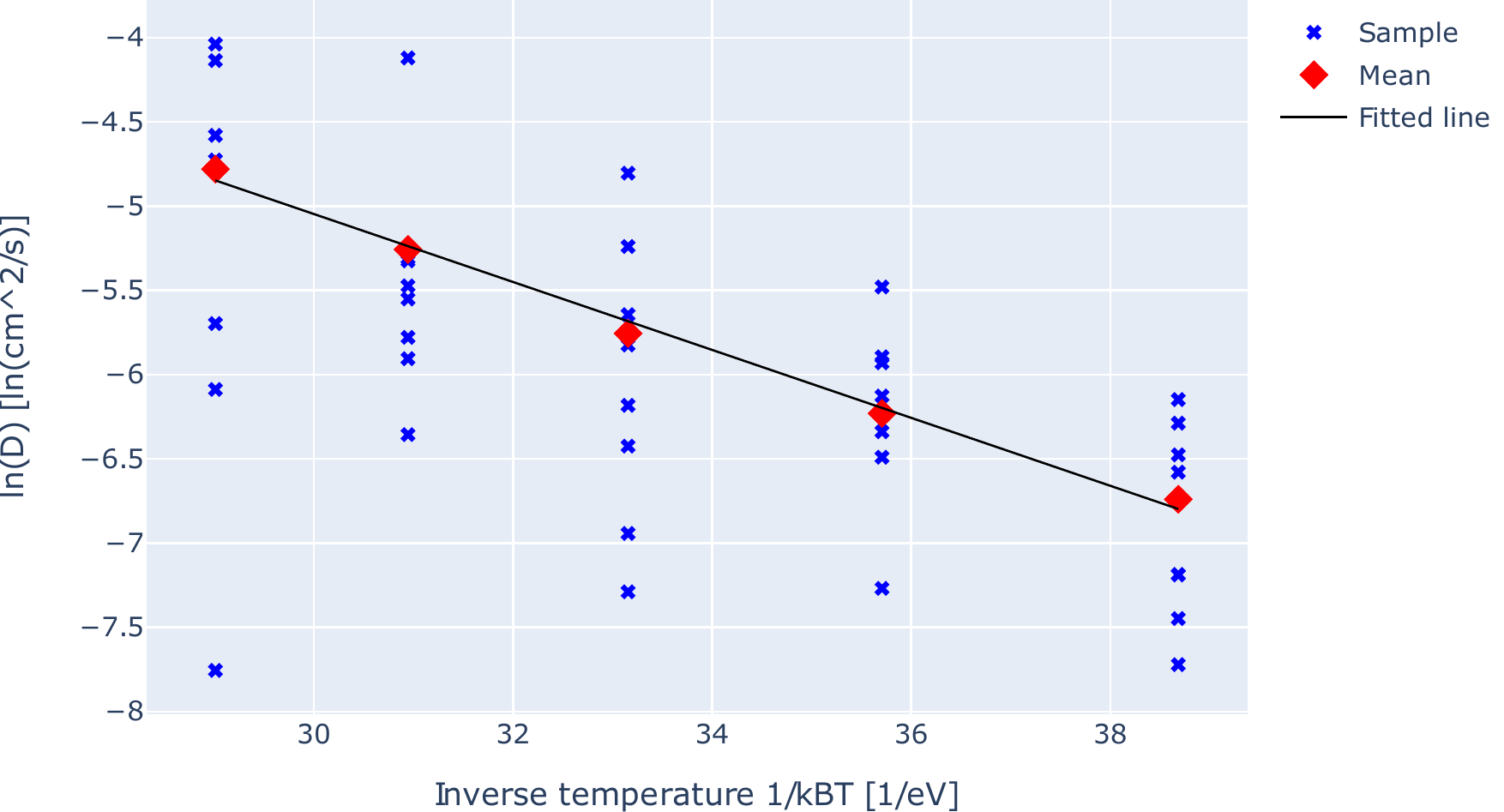}
\caption{Arrhenius plot of lithium diffusion coefficient. One trajectory at 325 K has a very small diffusion coefficient ($8.2\times {10}^{-7}$ cm${}^2$/s), which is not shown in this figure.}
\label{fig:li_arrhenius}
\end{figure*}

The trajectory of the lithium atom was sampled every 0.01 ps for 80 ps, starting 20 ps after the initial state of the MD simulation. The mean squared distance (MSD) of the lithium atom along the $[111]$ direction was calculated by changing the time span from 0.05 ps to 10 ps, and the diffusion coefficient was calculated by fitting the linear coefficient of the MSD with respect to the time span.\cite{doi:10.1021/acs.jctc.5b00574} The Arrhenius plot is shown in Fig. \ref{fig:li_arrhenius}. The fitted line was calculated from the series of the mean diffusion coefficients at each temperature. The activation energy corresponded to the coefficient of the fitted line, which was determined to be  0.202 eV for the activation energy of lithium diffusion.

\clearpage
\section*{Data 14: Computational details of MOF}
\begin{figure}[hbp]
\centering
\includegraphics[width=0.90\linewidth]{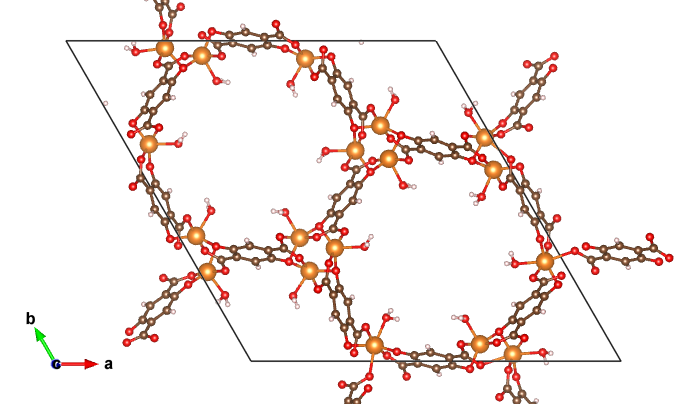}
\caption{Crystal structure of hydrated MOF-74-Mg viewed along the c-axis. Elements are represented by red (oxygen), orange (magnesium), gray (carbon), and white (hydrogen).}
\label{fig:mof}
\end{figure}

The crystalline structures of some representative MOFs were optimized using PFP. The starting crystalline structures were obtained from the Cambridge Structure Database (CSD).\cite{Groom:bm5086} The initial structures were cleaned by removing the physically adsorbed molecules in the pores of the MOFs. Water molecules that are chemically bound to the metal centers were maintained. We call these structures ``hydrated'' structures. Other minor cleansing procedures were performed by adding hydrogen atoms and removing overlapping atoms to ensure physically reasonable crystal structures and stoichiometries. Geometry optimization was performed on each MOF to determine the lowest energy configuration. The convergence criterion for the geometry optimization is for the maximum force on any atom to reach below 5 meV/\angstrom. The Broyden–Fletcher–Goldfarb–Shanno (BFGS) algorithm was used to optimize both the cell geometry and atom positions.\cite{Furukawa1230444}
Figure \ref{fig:mof} shows the crystal structure of hydrated MOF-74-Mg.

\clearpage
\section*{Data 15: Comparison of individual cell parameters of MOFs calculated by PFP}

\begin{table*}[hbp]
    \centering
    \caption{Unit cell parameters and volumes of the selected MOFs. Each experimental crystal structure was identified using the Cambridge Structure Database (CSD) identifier.\cite{Groom:bm5086} $\Delta \mathrm{V}_{\mathrm{exp}}$ represents the relative volumetric error of the PFP and PFP+D3 optimized geometries relative to the experimentally reported crystal structures.} \label{tab:mof}
    \begin{tabular}{cccllllllll}
    Metal & CSD itentifier & Data type & a (\angstrom) & b (\angstrom) & c (\angstrom) & $\alpha$ (deg.) & $\beta$ (deg.) & $\gamma$ (deg.) & V (\angstrom${}^3$) & $\Delta \mathrm{V}_{\mathrm{exp}}$ (\%) \\ \hline
MOF-5  & SAHYIK  & PFP  & 26.20  & 26.20  & 26.20  & 90.0  & 90.0  & 90.0  & 17982  & 6.32 \\
  &   & PFP+D3  & 26.15  & 26.15  & 26.15  & 90.0  & 90.0  & 90.0  & 17874  & 5.68 \\
  &   & Exp.\cite{doi:10.1126/science.1067208} & 25.67  & 25.67  & 25.67  & 90.0  & 90.0  & 90.0  & 16913  & \\
Cu-BTC  & FIQCEN  & PFP  & 26.50  & 26.47  & 26.47  & 90.0  & 89.9  & 90.1  & 18566  & 1.56 \\
  &   & PFP+D3  & 26.42  & 26.34  & 26.35  & 90.0  & 89.8  & 90.1  & 18337  & 0.30 \\
  &   & Exp.\cite{doi:10.1126/science.283.5405.1148}  & 26.34  & 26.34  & 26.34  & 90.0  & 90.0  & 90.0  & 18281  & \\
ZIF-90  & WOJGEI  & PFP  & 17.28  & 17.28  & 17.28  & 90.0  & 90.0  & 90.0  & 5161  & 0.17 \\
  &   & PFP+D3  & 17.13  & 17.13  & 17.13  & 90.0  & 90.0  & 90.0  & 5026  & -2.45 \\
  &   & Exp.\cite{doi:10.1021/ja805222x}  & 17.27  & 17.27  & 17.27  & 90.0  & 90.0  & 90.0  & 5152  & \\
MIL-53-Al  & SAVUN  & PFP  & 6.73  & 17.27  & 13.07  & 90.0  & 90.0  & 90.0  & 1517  & 7.47 \\
  &   & PFP+D3  & 6.70  & 17.21  & 13.04  & 90.0  & 90.0  & 90.0  & 1505  & 6.62 \\
  &   & Exp.\cite{https://doi.org/10.1002/chem.200305413}  & 6.61  & 16.68  & 12.81  & 90.0  & 90.0  & 90.0  & 1412  & \\
MOF-74-Mg  & VOGTIV  & PFP  & 26.20  & 26.21  & 7.05  & 90.0  & 90.0  & 120.0  & 4194  & 5.79 \\
  &   & PFP+D3  & 26.10  & 26.10  & 7.01  & 90.0  & 90.0  & 120.0  & 4132  & 4.23 \\
  &   & Exp.\cite{https://doi.org/10.1002/ejic.200701284}  & 26.03  & 26.03  & 6.76  & 90.0  & 90.0  & 120.0  & 3965  & \\
MOF-74-Co  & NAVJAW  & PFP  & 26.13  & 26.13  & 7.04  & 90.0  & 90.0  & 120.0  & 4164  & 4.97 \\
  &   & PFP+D3  & 26.01  & 26.01  & 6.94  & 90.0  & 90.0  & 120.0  & 4064  & 2.46 \\
  &   & Exp.\cite{https://doi.org/10.1002/anie.200501508}  & 26.11  & 26.11  & 6.72  & 90.0  & 90.0  & 120.0  & 3967  & \\
MOF-74-Ni  & LEJRIC  & PFP  & 26.05  & 26.04  & 6.97  & 90.0  & 90.0  & 120.0  & 4093  & 4.71 \\
  &   & PFP+D3  & 25.89  & 25.88  & 6.88  & 90.0  & 90.0  & 120.0  & 3993  & 2.16 \\
  &   & Exp.\cite{B515434K}  & 25.98  & 25.98  & 6.69  & 90.0  & 90.0  & 120.0  & 3909  & \\
MOF-74-Zn  & WOBHEB  & PFP  & 26.24  & 26.24  & 7.00  & 90.0  & 90.0  & 120.0  & 4176  & 4.87 \\
  &   & PFP+D3  & 26.11  & 26.11  & 6.95  & 90.0  & 90.0  & 120.0  & 4103  & 3.03 \\
  &   & Exp.\cite{https://doi.org/10.1002/chem.200701370}  & 26.26  & 26.26  & 6.67  & 90.0  & 90.0  & 120.0  & 3983  & \\
    \end{tabular}
\end{table*}

\clearpage
\section*{Data 16: Metropolis sampling method}
The MC moves used in the Cu-Au alloy order-disorder transition simulation were conducted using Metropolis sampling.\cite{Understanding_Molecular_Simulation}, such that an arbitrary pair of atoms is swapped and the structure is relaxed, and the energy change ($\Delta E$) is recorded. Since the optimization of the structure is computationally expensive, it is performed only at every 100 steps. Then, the Boltzmann factor, $\exp\left(-\Delta E/k_BT\right)$, is computed and compared with a randomly selected number between 0 and 1 according to a uniform distribution used to determine whether the move should be accepted or rejected. If the random number is smaller than the Boltzmann factor, the move is accepted; otherwise, the move is rejected. The MC loop was iterated over 200,000 steps to ensure equilibrium. The final structure of the MC runs was characterized based on the Voronoi-weighted Steinhardt parameters.\cite{doi:10.1063/1.4774084} These parameters are suitable for characterizing the ordering of the atomic arrangement.
To calculate under periodic conditions, a 10 \angstrom \ vacuum was applied along each axis. To reduce the computational cost, a 7 \angstrom \ vacuum was also used.

\clearpage
\section*{Data 17: High-temperature multi-element dataset description}
In this study, we provide an atomic structure dataset called the high-temperature multi-element 2021 (HME21) dataset, which consists of a portion of the PFP dataset (Supplementary Script 2).

There are several reasons for providing HME21.
The first reason is enabling readers to further analyze the properties of PFP by providing a portion of the PFP dataset. We believe that the nature of the dataset, such as the number of element types in a single structure, or the variety of neighboring atoms, plays an important role in achieving the molecular dynamics simulation results shown in this study.
The second reason is to be a standard benchmark for future universal NNP development. As mentioned in the Introduction, existing datasets were generated based on known structures, such as the molecule or crystal structures. During our exploration while building the new dataset, we noticed that the existing NNP architectures did not behave as expected when the dataset components were far from stable atomic structures. In most cases, the hyperparameters must be retuned to proceed with the training process. We believe that the HME21 dataset will serve as a guide for the design of future universal NNPs.
The third reason is to provide an NNP architecture benchmark combined with the dataset, allowing readers to confirm the benchmark results themselves. The corresponding benchmark results of the NNP-architecture are presented in Supplementary Data 18.

HME21 corresponds to the disordered dataset described in Supplementary Data 10. We regarded this subcomponent as a typical structure in the PFP dataset. It contains multiple elements in a single structure and was sampled through a high-temperature molecular dynamics simulation. Thus, the structures are far from stable and contain less prior specific domain knowledge, such as the molecule or crystal structures.
The structures were sampled from the entire disordered dataset and randomly split into training, validation, and test sub-datasets at a ratio of 8:1:1. The numbers of structures in these datasets are 19956, 2498, and 2495, respectively.
There are a total of 37 elements in the HME21 dataset, i.e., H, Li, C, N, O, F, Na, Mg, Al, Si, P, S, Cl, K, Ca, Sc, Ti, V, Cr, Mn, Fe, Co, Ni, Cu, Zn, Mo, Ru, Rh, Pd, Ag, In, Sn, Ba, Ir, Pt, Au, and Pb. Each dataset contains structural information (element types, atomic positions, and cell shapes) and target values (energy and atomic forces). All structures are under periodic boundary conditions. The energy is shifted such that the energy of a single atom located in a vacuum becomes zero. The length is in ångströms ($10^{-10}$ m), and the energy is in electronvolts (eV).

The details of the DFT calculation conditions are provided in Supplementary Data 10.

\clearpage
\section*{Data 18: Neural network architecture benchmark using HME21}
To show the performances of NNP architectures for multi-element structures that are far from having a stable state, we applied a benchmark using the HME21 dataset (see Supplementary Data 17).
For this benchmark, we selected TeaNet\cite{teanet}, SchNet~\cite{schnet2017}, PaiNN~\cite{painn}, and NequIP~\cite{nequip}.
TeaNet corresponds to the base model of PFP. It treats tensor representations as higher-order geometric features.
SchNet uses the bond length for spatial information and employs a convolution with rotationally invariant filters. 
This has been well examined using various datasets, and its limited representation power has been discussed. 
PaiNN incorporates a vector representation to resolve the problem of a limited representation of rotationally invariant filters of SchNet.
On the other hand, NequIP uses spherical harmonics-based representations. 
The experimental code for both SchNet and PaiNN is based on the repository found at \mbox{https://github.com/learningmatter-mit/NeuralForceField}, whereas the experimental code for NequiP is based on the repository found at \mbox{https://github.com/mir-group/nequip}.

Next, we discuss the choice of hyperparameter. 
To optimize the performance with respect to the validation set, the hyperparameter selection procedure is based on a grid search and manual hyperparameter tuning.
For TeaNet, we use a four-layer model. We first set the energy loss coefficient $c_{le}$ (energy per atom MSE) to $0.0001$ and retrained it using $c_{le} = 1.0$ and $c_{le} = 10.0$ , whereas the force loss coefficient $c_{lf}$ remained constant at $1.0$. The batch size was set to $16$, and the learning rate was initialized to $0.001$.
For SchNet, we use a four-layer model, where the energy loss coefficient was set to $0.05$, the batch size was set to $32$, and the learning rate was initialized to $0.0005$.
For PaiNN, we use a three-layer model, where the energy loss coefficient was set to $0.05$, the batch size was set to $32$, and the learning rate was initialized to $0.0005$.
For NequIP, we use a five-layer model with different maximum rotation orders $l_\mathrm{max} \in \{0, 1, 2\}$.
For the five-layer model, the energy loss coefficient was set to $0.01$ and the learning rate was initialized to $0.001$. 
For $l_\mathrm{max} \in \{0,1\}$, we found that setting the batch size to $32$ worked best, whereas for $l_\mathrm{max} = 2$, setting the batch size to $64$ was preferable.
We set the cutoff distance to $6.0$ \angstrom for all architectures.

The results are presented in Table \ref{tab:nnp_benchmarks}. TeaNet performed well in terms of both the energy and force metrics, which indicates that the TeaNet architecture is suitable for multielement structures which are far from stable coordination. 

The results, including the implementation of the TeaNet architecture, are available along with those of the HME21 dataset (Supplementary Script 3).

\begin{table*}[hbp]
  \centering
  \begin{tabular}{lD{.}{.}{3}D{.}{.}{4}D{.}{.}{4}}
 & \multicolumn{1}{c}{Energy MAE} & \multicolumn{1}{c}{Force MAE}  & \multicolumn{1}{c}{Force XYZ MAE} \\
Architecture & \multicolumn{1}{c}{[meV/atom]} & \multicolumn{1}{c}{[eV/\angstrom]}  & \multicolumn{1}{c}{[eV/\angstrom]} \\ \hline
TeaNet (PFP base model) & \multicolumn{1}{B{.}{.}{3}}{19.6} & \multicolumn{1}{B{.}{.}{3}}{0.174} & \multicolumn{1}{B{.}{.}{3}}{0.153} \\
SchNet & 33.6 & 0.283 & 0.247 \\
PaiNN & 22.9 & 0.237 & 0.208 \\
NequIP ($l_\mathrm{max}=0$) & 52.2 & 0.249 & 0.225 \\
NequIP ($l_\mathrm{max}=1$) & 53.3 & 0.233 & 0.206 \\
NequIP ($l_\mathrm{max}=2$) & 47.8 & 0.199 & 0.175 \\
  \end{tabular}
  \caption{Benchmark performance of NNP for the force and energy prediction for the HME21 dataset.
  ``Energy MAE'' corresponds to the mean absolute error of the energies of structures divided by their numbers of atoms, ``Force MAE'' corresponds to the mean absolute error of the norm of force vectors, and ``Force XYZ MAE'' corresponds to the mean absolute error of the force vector component.
  } 
  \label{tab:nnp_benchmarks}
\end{table*}

\clearpage
\section*{Data 19: Statistical information of PFP dataset}
Table \ref{tab:pfp_dataset_info} lists the statistical information of the PFP molecule and PFP crystal datasets.

\begin{table*}[hbp]
    \centering
    \caption{Statistical information of (a) PFP molecule and (b) PFP crystal datasets. Here, ``\# of atoms'' and ``\# of elements'' correspond to the numbers of atoms and elements in each structure, respectively. In addition, ``Energy,'' corresponding to the energy of the structures divided by their number of atoms, and the energy of single atoms in a vacuum, is set to zero. ``Force'' corresponds to the $L^2$ norm of the force vector of each atom. The range, mean, and standard deviation (SD) are shown.}
    \label{tab:pfp_dataset_info}

    (a) PFP molecule dataset\\
    \begin{tabular}{l|c|ccc|ccc|cc|cc}
    {} & \# of & \multicolumn{3}{c|}{\# of atoms} & \multicolumn{3}{c|}{\# of elements} & \multicolumn{2}{c|}{Energy [eV/atom]} & \multicolumn{2}{c}{Force [eV/$\angstrom$]} \\
    Structure & Structure & Range & Mean & SD & Range & Mean & SD & Mean & SD & Mean & SD \\
    \hline
    Optimize  &  2.6$\times10^{5}$  &  2 -- 26  &  15  &  3.4  &  1 -- 9  &  4.1  &  1.0  &  -4.1  &  0.35  &  1.4$\times10^{-5}$  &  5.8$\times10^{-3}$  \\
    NMS  &  4.6$\times10^{6}$  &  2 -- 26  &  15  &  3.6  &  1 -- 9  &  3.8  &  0.86  &  -4.0  &  0.35  &  3.8  &  6.7  \\
    MD  &  7.8$\times10^{5}$  &  2 -- 26  &  16  &  3.6  &  1 -- 6  &  3.7  &  0.74  &  -4.0  &  0.32  &  7.7  &  15  \\
    MD with two molecules  &  4.4$\times10^{5}$  &  4 -- 40  &  23  &  5.2  &  1 -- 6  &  4.5  &  0.80  &  -3.9  &  0.31  &  3.8  &  12  \\
    \end{tabular}

    (b) PFP Crystal Dataset \\
    \begin{tabular}{l|c|ccc|ccc|cc|cc}
    {} & \# of & \multicolumn{3}{c|}{\# of atoms} & \multicolumn{3}{c|}{\# of elements} & \multicolumn{2}{c|}{Energy [eV/atom]} & \multicolumn{2}{c}{Force [eV/\angstrom]}, \\
    Structure & Structure & Range & Mean & SD & Range & Mean & SD & Mean & SD & Mean & SD \\
    \hline
    Molecule  &  1.0$\times10^{6}$  &  2 -- 36  &  13  &  5.3  &  1 -- 9  &  4.0  &  1.2  &  -2.8  &  1.2  &  2.6  &  7.1  \\
    Bulk, cell compression / expansion  &  5.9$\times10^{4}$  &  2 -- 16  &  5.5  &  4.2  &  1 -- 2  &  1.9  &  0.28  &  -3.1  &  1.3  &  0.91  &  5.5  \\
    Bulk, site position displacement  &  3.2$\times10^{4}$  &  4 -- 112  &  15  &  7.7  &  1 -- 2  &  1.9  &  0.25  &  -2.9  &  1.3  &  15  &  39  \\
    Bulk, cell shear deformation  &  4.4$\times10^{4}$  &  2 -- 16  &  5.3  &  4.0  &  1 -- 2  &  2.0  &  0.19  &  -3.2  &  1.3  &  0.18  &  0.60  \\
    Bulk, cell tensile deformation  &  6.0$\times10^{4}$  &  2 -- 16  &  5.7  &  4.2  &  1 -- 2  &  2.0  &  0.16  &  -3.2  &  1.3  &  0.61  &  3.4  \\
    Cluster  &  2.2$\times10^{4}$  &  13 -- 13  &  13  &  0.0  &  1 -- 1  &  1.0  &  0.0  &  -1.7  &  1.2  &  11  &  33  \\
    Disordered  &  8.1$\times10^{5}$  &  2 -- 128  &  27  &  19  &  1 -- 28  &  7.8  &  6.2  &  -2.5  &  0.99  &  11  &  26  \\
    Slab  &  7.4$\times10^{5}$  &  3 -- 192  &  18  &  10  &  1 -- 2  &  1.9  &  0.24  &  -2.9  &  1.3  &  1.3  &  5.8  \\
    Adsorption, geometrical optimization  &  9.7$\times10^{4}$  &  6 -- 121  &  33  &  19  &  2 -- 8  &  4.3  &  0.86  &  -3.5  &  0.91  &  0.76  &  8.2  \\    
    Adsorption, random placement  &  2.8$\times10^{5}$  &  5 -- 400  &  52  &  25  &  1 -- 6  &  4.5  &  0.78  &  -3.1  &  0.93  &  2.9  &  10  \\
    \end{tabular}
\end{table*}

\clearpage
\section*{Data 20: PFP molecule mode with out-of-domain elements}
\begin{figure}[hbp]
\centering
\includegraphics[width=0.70\linewidth]{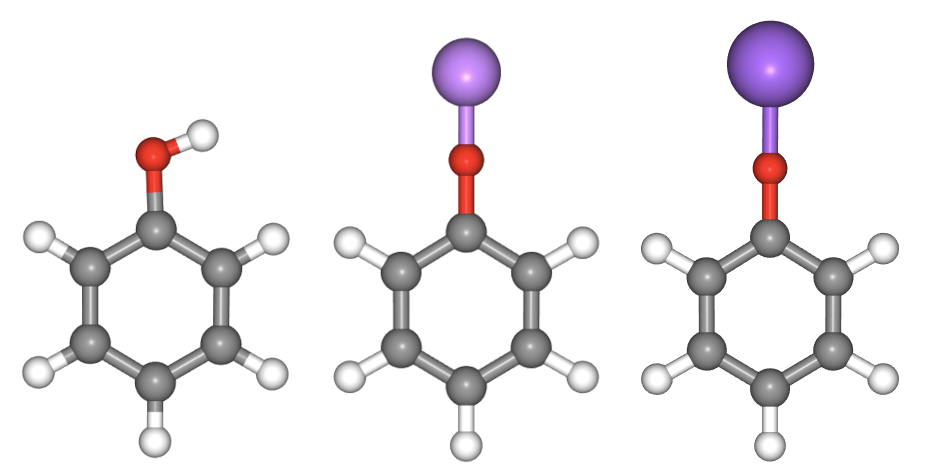}
\caption{Optimized structures of phenol, lithium phenoxide, and sodium phenoxide using PFP with crystal mode.}
\label{fig:phenols}
\end{figure}

In this section, the domain transferability of the PFP under different DFT conditions is examined. We compared the molecule structures of phenol, lithium phenoxide, and sodium phenoxide estimated using both PFP with molecule mode and PFP with crystal mode. In crystal mode, all elements are included in the dataset, and it is expected that all three of these molecules are in-domain molecules. Figure \ref{fig:phenols} shows the optimized structures obtained using crystal mode. By contrast, in molecule mode, C, H, and O are included in the dataset but Li and Na are not. Therefore, if an NNP is trained using a molecule dataset only, it is clear that there is no information at all regarding Li and Na elements, and the structural estimations of lithium phenoxide and sodium phenoxide will fail.
However, PFP was  concurrently trained using both the molecule and crystal datasets. This means that, even under molecule mode inference, PFP knows the behaviors of Li and Na atoms in crystal mode, and there is a possibility of transferring them to molecule mode inference.

Comparisons of the bond and angle parameters of the optimized structures of these three molecules between the crystal and molecule modes are shown in Table \ref{tab:phenol_parameters}. First, the positions of the Li and Na atoms were consistent in both the crystal and molecule modes. It can be said that the behaviors of Li and Na atoms under molecule mode imitate those of crystal mode. Second, the ratios of the bond distances in the benzene ring (C--C, C--H, C--O) between the crystal and molecule modes were mostly constant among the three molecules. Owing to the difference in DFT conditions, the bond distances of phenol are slightly different between the molecule and crystal modes. It can be interpreted that the benzene rings of lithium phenoxide and sodium phenoxide follow the molecule dataset.

\begin{table*}[hbp]
  \centering
  \begin{tabular}{l|D{.}{.}{3}D{.}{.}{3}D{.}{.}{3}|D{.}{.}{3}D{.}{.}{3}D{.}{.}{3}|D{.}{.}{3}D{.}{.}{3}D{.}{.}{3}}
          & \multicolumn{3}{c|}{Crystal mode} & \multicolumn{3}{c|}{Molecule mode} & \multicolumn{3}{c}{Ratio} \\
          & \multicolumn{1}{c}{H} & \multicolumn{1}{c}{Li} & \multicolumn{1}{c|}{Na}& \multicolumn{1}{c}{H} & \multicolumn{1}{c}{Li} & \multicolumn{1}{c|}{Na}& \multicolumn{1}{c}{H} & \multicolumn{1}{c}{Li} & \multicolumn{1}{c}{Na} \\ \hline
C--C bond min [\angstrom]    &   1.396 &  1.396 &  1.396 &  1.390 &  1.391 &  1.390 &  1.004 &  1.004 &  1.005 \\
C--C bond max [\angstrom]    &   1.402 &  1.416 &  1.423 &  1.398 &  1.413 &  1.419 &  1.003 &  1.002 &  1.003 \\
C--H bond min [\angstrom]    &   1.091 &  1.092 &  1.092 &  1.086 &  1.087 &  1.087 &  1.005 &  1.005 &  1.005 \\
C--H bond max [\angstrom]    &   1.094 &  1.093 &  1.094 &  1.089 &  1.087 &  1.088 &  1.005 &  1.006 &  1.006 \\
C--O bond [\angstrom]        &   1.379 &  1.336 &  1.323 &  1.362 &  1.313 &  1.303 &  1.012 &  1.017 &  1.015 \\
O--X bond [\angstrom]        &   0.973 &  1.618 &  1.980 &  0.963 &  1.618 &  2.016 &  1.010 &  1.000 &  0.982 \\
C--O--X angle [degree]    & 109.395 &179.996 &179.645 &109.574 &179.997 &179.610 &  0.998 &  1.000 &  1.000
  \end{tabular}
  \caption{Bond distances and angles of optimized structures of phenol, lithium phenoxide, and sodium phenoxide estimated using PFP with crystal mode and PFP with molecule mode. The ``H'', ``Li'', and ``Na'' columns correspond to phenol, lithium phenoxide, and sodium phenoxide, respectively. In addition, ``bond min'' and ``bond max'' correspond to the minimum and maximum bond distances of the corresponding bonds, respectively; and ``X'' corresponds to the H, Li, and Na atoms in the hydroxy group. Finally, the ``Ratio'' column shows the ratio of values between crystal and molecule modes.}
  \label{tab:phenol_parameters}
\end{table*}

\clearpage
\bibliographystyle{naturemag}
\bibliography{references}